\documentclass[AMA,Times1COL]{WileyNJDv5} 
\usepackage{siunitx}
\usepackage{subfigure}
\usepackage{url}
\usepackage{float}
\usepackage{xcolor}
\usepackage{tabularx}
\usepackage{makecell}
\usepackage{multirow}
\usepackage{bbding}
\usepackage{colortbl}
\usepackage{graphicx}
\definecolor{skyblue}{RGB}{0, 128, 200}
\definecolor{skygreen}{RGB}{0, 116, 23}
\definecolor{skyred}{RGB}{178, 127, 31}
\usepackage{multirow}

\articletype{Article Type}%

\received{Date Month Year}
\revised{Date Month Year}
\accepted{Date Month Year}
\journal{Journal}
\volume{00}
\copyyear{2023}
\startpage{1}

\begin{document}

\title{Joint$\lambda$: Orchestrating Serverless Workflows \\ on Jointcloud FaaS Systems}

\author[1]{Rui Li$^{*}$}
\author[1]{Jianfei Liu$^{*}$}
\author[1]{Zhilin Yang}
\author[1]{Peichang Shi}
\author[2]{Guodong Yi}
\author[1]{Huaimin Wang}
\authormark{Li \textsc{et al.}}
\titlemark{Joint$\lambda$}

\address[1]{State Key Laboratory of Complex \& Critical Software Environment, College of Computer Science and Technology, National University of Defense Technology, Changsha, Hunan, China}

\address[2]{Xiangjiang Lab, Hunan University Of Technology and Business, Changsha, Hunan,  China}

\corres{Huaimin Wang, National Key Laboratory of Parallel and Distributed Computing, National University of Defense Technology,  Changsha, Hunan,  China. \email{whm\_w@163.com}}


\abstract[Abstract]{Existing serverless workflow orchestration systems are predominantly designed for a single-cloud FaaS system, leading to vendor lock-in. This restricts performance optimization, cost reduction, and availability of applications. However, orchestrating serverless workflows on Jointcloud FaaS systems faces two main challenges: (1) additional overhead caused by centralized cross-cloud orchestration; and (2) a lack of reliable failover and fault-tolerant mechanisms for cross-cloud serverless workflows.

To address these challenges, we propose Joint$\lambda$, a distributed runtime system designed to orchestrate serverless workflows on multiple FaaS systems without relying on a centralized orchestrator. Joint$\lambda$ introduces a compatibility layer, Backend-Shim, leveraging inter-cloud heterogeneity to optimize makespan and reduce costs with on-demand billing. By using function-side orchestration instead of centralized nodes, it enables independent function invocations and data transfers, reducing cross-cloud communication overhead. For high availability, it ensures exactly-once execution via datastores and failover mechanisms for serverless workflows on Jointcloud FaaS systems. We validate Joint$\lambda$ on two heterogeneous FaaS systems, AWS and Aliyun, with four workflows. Compared to the most advanced commercial orchestration services for single-cloud serverless workflows, Joint$\lambda$ reduces makespan by up to 3.3$\times$ while saving up to 65\% in cost. Joint$\lambda$ is also up to 4.0$\times$ faster than state-of-the-art orchestrators for cross-cloud serverless workflows, while achieving competitive cost in representative scenarios and providing strong execution guarantees.}

\keywords{Cloud computing, Serverless Computing, Jointcloud Computing, Function-as-a-service, Workflows}

\maketitle

\footnotetext[1]{Rui Li and Jianfei Liu contributed equally to this work and share first authorship.}

\renewcommand\thefootnote{}

\renewcommand\thefootnote{\fnsymbol{footnote}}
\setcounter{footnote}{1}

\section{Introduction}
\label{Introduction}


Serverless computing simplifies cloud programming \cite{jonas2019cloud}, separates application development and infrastructure management (e.g., auto-scaling), enables fine-grained on-demand billing, and charges users only for the resources they actually use. Serverless computing provides two simple but efficient high-level abstractions: stateless basic computing units (Function as a Service, FaaS) and scalable backend services (Backend as a Service, BaaS) such as data storage. Developers decompose applications into a series of logically independent functions, using external storage to transfer intermediate data. However, the fine-grained division also creates complex workflows for FaaS applications. Users typically need to logically define workflows (e.g., Directed Acyclic Graphs, DAGs), and then use an orchestrator to organize and coordinate the dispersed functions to complete the defined workflow.



There are many implementations of serverless workflow orchestration systems \cite{fouladi2017encoding,fouladi2019laptop,zhang2020fault,liu2023doing,li2022faasflow,burckhardt2021durable,ao2018sprocket,malawski2020serverless,yu2023following} nowadays.
However, almost all serverless workflow orchestration systems are designed to work on a single FaaS system, leading to vendor lock-in, potentially compromising the performance, cost, and availability of applications. The serverless workflows on a single FaaS platform face several limitations: 1) \textbf{The lack of support for heterogeneous accelerators.} There have been many works \cite{wen2023characterizing,copik2021sebs,maissen2020faasdom,yu2020characterizing,kim2019practical} exposing the performance differences among multiple homogeneous FaaS systems (e.g., CPU-only support). However, few FaaS systems \cite{ali,yang2022infless,du2022serverless} currently support heterogeneous accelerators, which are crucial for accelerating computation. 
2)  \textbf{Different FaaS systems incur varying costs.} The FaaS system uses a pay-as-you-go billing model. Functions running on different FaaS systems generate significantly different charges. External serverless workflow orchestrators often charge based on state transitions rather than actual resource consumption, leading to unfair billing. 3)  \textbf{Availability of cloud services.} Strict regulations on data and operational sovereignty (e.g., the General Data Protection Regulation, GDPR) dictate where data can be stored and where jobs can be executed. Not all cloud providers have data centers in every country, making it challenging to comply with these regulations \cite{yang2023skypilot}. Additionally, cloud customers suffer significant losses from reduced availability during regional or cloud-wide failures \cite{googleoutage}. 





We believe that future FaaS applications will be built on multiple clouds, such as Jointcloud computing \cite{shi2019inherent} and Sky computing \cite{keahey2009sky}, which facilitate workload deployment and migration across clouds as well as peer-to-peer collaboration between multiple cloud service providers (CSPs). However, orchestrating serverless workflows on Jointcloud FaaS systems faces two main challenges: 1) \textbf{Additional overhead caused by centralized cross-cloud orchestration.} Functions are short-lived, so state transitions at the orchestrator occur frequently \cite{li2022faasflow}. Each centralized state transition adds at least one additional communication hop, leading to non-negligible latency overhead. Moreover, a centralized cross-cloud orchestrator incurs additional cost due to long-running orchestration nodes (e.g., virtual machines, VMs) and frequent cross-cloud data transfers. 2) \textbf{A lack of reliable failover and fault-tolerant mechanisms.} Function execution and invocation may fail for many reasons in multi-cloud environments, e.g., cloud failures. FaaS systems typically provide only weak at-least-once execution guarantees (e.g., retries), which may cause repeated executions and inconsistent workflow outputs. Without reliable failover and exactly-once execution semantics for cross-FaaS workflows, it is difficult to ensure workflow reliability.





To overcome single-cloud limitations and address these challenges, we present Joint$\lambda$, a distributed runtime system for orchestrating serverless workflows on multiple FaaS systems. Without a centralized orchestrator, Joint$\lambda$ located with each user function as an additional runtime library (function-side workflow orchestrator), enables the control flow from function to function. This can accelerate makespan and reduce costs via inter-cloud heterogeneity, ensure high availability with fault tolerance and failover, and avoid extra communication during state transitions. To achieve this, we prototyped the Joint$\lambda$ upon Unum \cite{liu2023doing} which is an application-level distributed FaaS workflow orchestrator based on a single cloud. 


 In Joint$\lambda$, we introduce a distributed compatibility layer, Backend-Shim, implemented based on the APIs of different FaaS systems and data storages, providing a standard interface for the function-side workflow orchestrator. Joint$\lambda$ manages data reads/writes and function invocations across various backend systems via the Backend-Shim. We also extend the intermediate representation of workflows to express complex invocation primitives and transfer primitives using function meta-information (sub-graph). We introduce a wrapper to manage function node entry/export, parsing cloud event objects (Joint$\lambda$Object) and transferring data efficiently. The function-side workflow orchestrator inside the wrapper implements current function orchestration using function invocation and data transfer primitives provided by the sub-graph. For high availability in unstable environments, we implement fault-tolerant and failover mechanisms. We also employ the majority rule principle for data storage location and a bitmap for cross-platform function collaboration, defining unique key naming for function reusability.

The main contributions of this paper are as follows.

 1)  We present Joint$\lambda$, a distributed runtime system for orchestrating serverless workflows on multiple FaaS systems, leveraging inter-cloud heterogeneity. It uses function-side distributed orchestration instead of centralized nodes to reduce cross-cloud communication overhead.

2) We achieve exactly-once execution guarantees and failover mechanisms to ensure high availability for serverless workflows on Jointcloud FaaS.

3) We validate Joint$\lambda$ on two heterogeneous FaaS systems, Amazon Web Services (AWS) Lambda \cite{awslambda} and Alibaba Cloud (Aliyun) Function Compute (FC) \cite{aliyunfc}, with four workflows. Compared with AWS Step Functions and Aliyun CloudFlow, our evaluation shows that Joint$\lambda$ reduces makespan by up to 3.3$\times$ and saves up to 65\% in cost. Joint$\lambda$ is also up to 4.0$\times$ faster than state-of-the-art orchestrators for cross-cloud serverless workflows with competitive cost in representative scenarios. Joint$\lambda$ further supports reliable failover for serverless workflows, incurring negligible additional overhead relative to the overall runtime and cost.




\section{Background and Motivation}





In serverless computing, developers write and upload functions to a FaaS platform, and these functions can be implemented in multiple high-level languages. A function responds to request events, including HTTP requests and BaaS events. When the function is invoked, the CSP configures a sandbox (e.g., a container \cite{docker} or a virtual machine \cite{agache2020firecracker,gviosr}), deploys the user code, and routes the request. FaaS platforms can quickly absorb bursty workloads, relieving developers from low-level operations and infrastructure management. In this section, we introduce the advantages of serverless workflows on Jointcloud FaaS systems and the limitations of existing cross-cloud FaaS workflow orchestrators.

\subsection{Why Serverless Workflows Can Benefit From Jointcloud FaaS Systems?}

\noindent  \emph{\textbf{Observation 1:} Existing FaaS systems vary in performance, with most lacking support for heterogeneous accelerators. Utilizing inter-cloud heterogeneity in workflows can significantly reduce the makespan.}


First, there have been many works \cite{wen2023characterizing,copik2021sebs,maissen2020faasdom,yu2020characterizing,kim2019practical} exposing the performance differences among multiple homogeneous FaaS systems (e.g., CPU-only support). Moreover, many important applications rely on heterogeneous accelerators (e.g., GPUs) to accelerate computation. However, very few FaaS systems \cite{ali,yang2022infless,du2022serverless} currently support heterogeneous accelerators. This limitation will be a serious obstacle for FaaS applications seeking to optimize makespan. In addition, workflow orchestrators supported by FaaS systems (e.g., orchestration services provided by major cloud service providers, CSPs) exhibit large performance differences when orchestrating multiple patterns \cite{wen2021measurement}. These limitations collectively constrain the overall performance of single-cloud FaaS workflows.
\begin{figure}[htbp]
\centering
\subfigure[P95 latency under batching of 2]{
\includegraphics[width=0.45\textwidth]{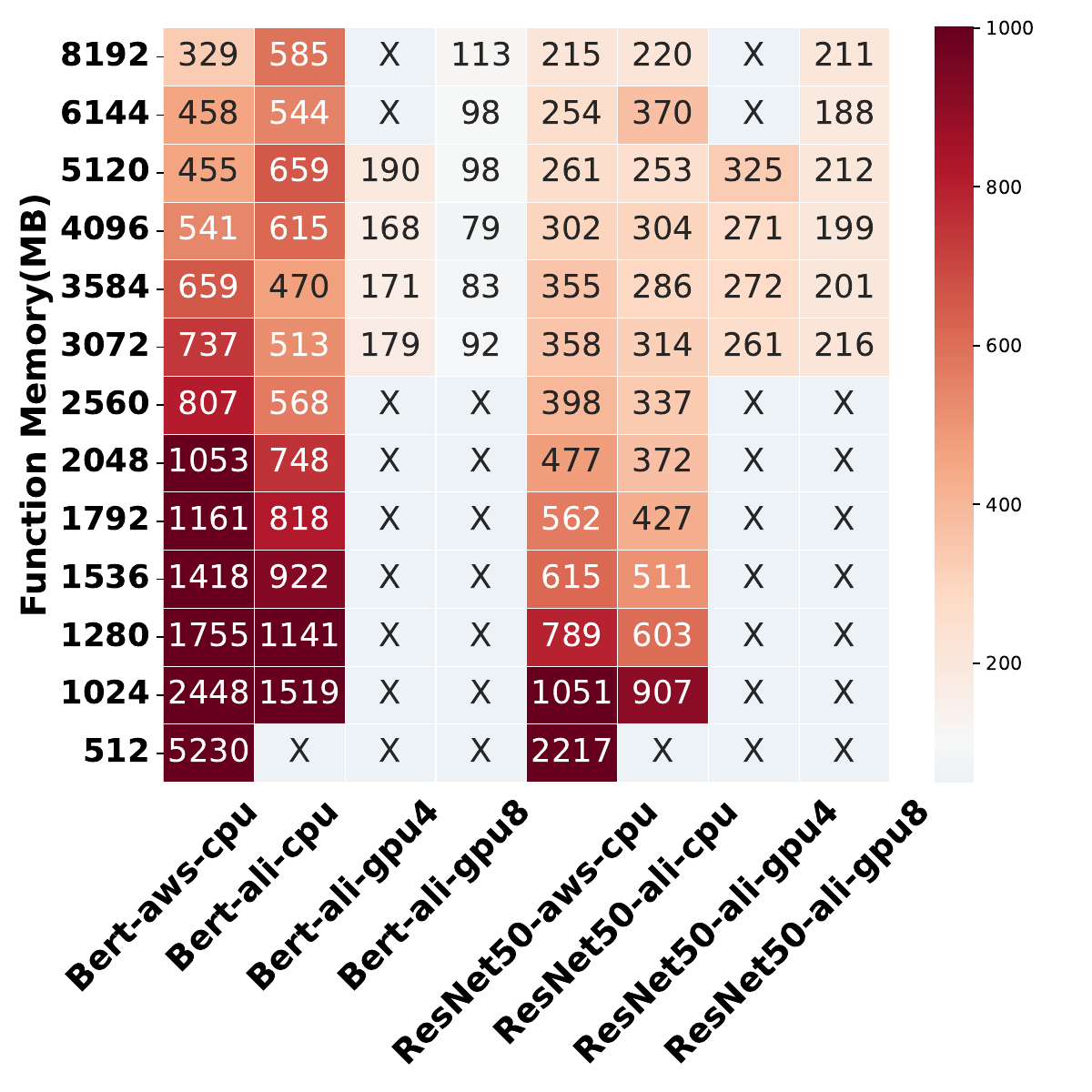} \label{batch2} 
}
\hfill
\subfigure[P95 latency under batching of 4]{
\includegraphics[width=0.45\textwidth]{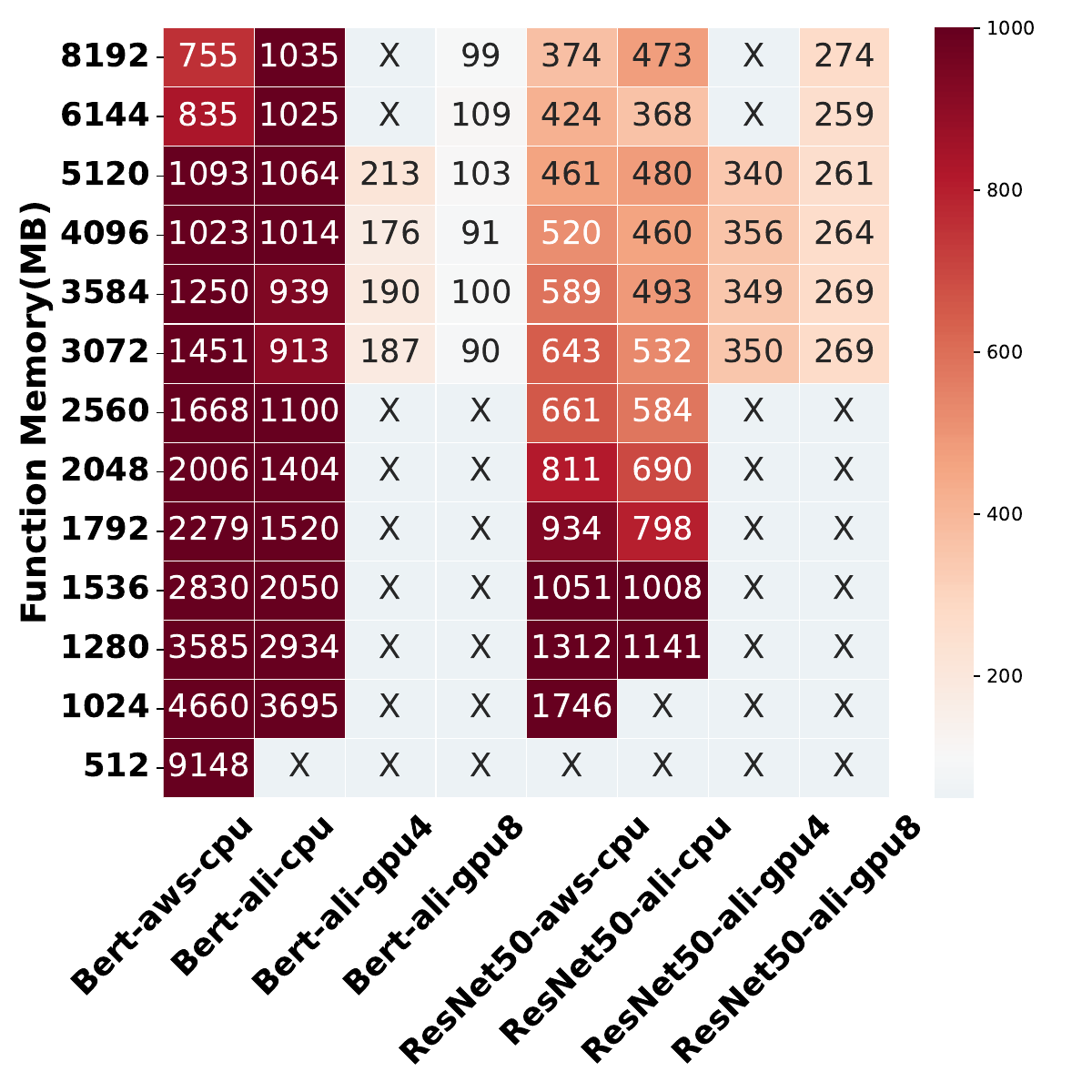} \label{batch4} 
}
\caption{P95 Inference Latency across Heterogeneous FaaS Configurations with Varying Memory Allocations. Note: The suffixes -aws-cpu, -ali-cpu, -ali-gpu4, and -ali-gpu8 denote AWS Lambda (CPU) and Aliyun Function Compute (CPU, 4GB A10 GPU, and 8GB A10 GPU) runtimes, respectively. 'X' indicates execution failure (e.g., OOM).}
\label{batch_l}
\end{figure}

The serverless architecture provides function abstraction to the user while hiding the specific details of resource management, e.g., the developer only knows the configured memory size but not the vCPU, network, and other relevant runtime data. This "function black box" results in underlying performance differences at runtime. Applications such as deep learning rely on hardware accelerators to accelerate computation. Figure~\ref{batch_l} presents the P95 latency distribution for BERT \cite{devlin2019bert} and ResNet50 \cite{he2016deep}. We evaluate these models across diverse FaaS runtimes, including mainstream CPU instances (Bert-aws-cpu, Bert-ali-cpu) and GPU-accelerated instances (Bert-ali-gpu4/8). The inference of BERT achieved up to a 7$\times$ and 15$\times$ acceleration on GPU-enabled FaaS systems compared to CPU-only FaaS systems for batch sizes of 2 and 4, respectively. The performance gap will be more pronounced with increasing model size or batch size.


\noindent  \emph{\textbf{Observation 2:} The FaaS model charges on demand based on GB*s, with total workflow cost positively correlated with total runtime. Leveraging inter-cloud heterogeneity, workflows can reduce execution costs.}



The FaaS system employs a pay-as-you-go billing model, where the total cost comprises the execution cost and the invocation cost. Execution cost is determined by the GB*s model (allocated resources multiplied by runtime) and depends on the resource unit price, allocated resources, and runtime. As a result, functions running on different FaaS systems generate significantly different charges.

\begin{figure}[htbp]
\centering
\includegraphics[width=0.6\textwidth]{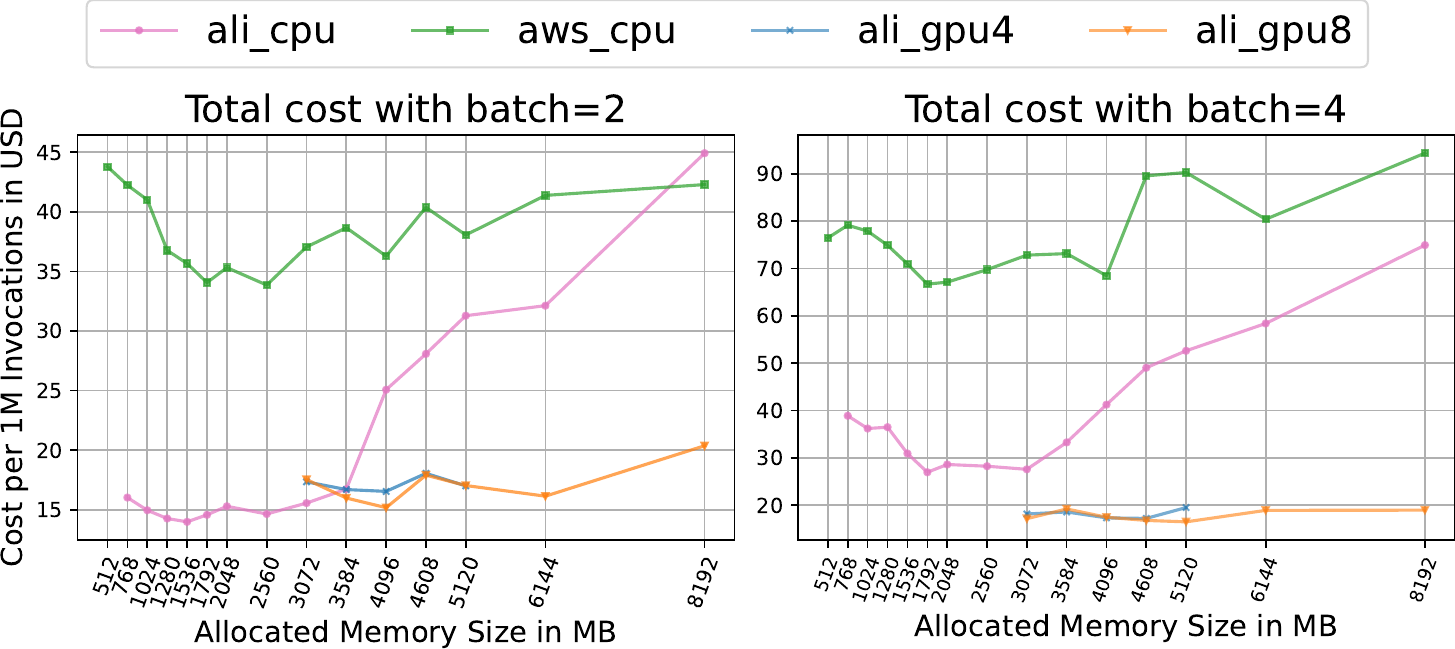} 
\caption{Average total cost per 1M invocations for BERT inference across heterogeneous FaaS configurations with Batch Size 2 and 4. Note: aws\_cpu, ali\_cpu, ali\_gpu4, and ali\_gpu8 denote AWS Lambda (CPU) and Aliyun Function Compute (CPU, 4GB A10 GPU, and 8GB A10 GPU) runtimes, respectively. The cost is measured in USD based on the memory-time resource pricing model of each provider.}
\label{batch_c}
\end{figure}

Figures \ref{batch_c} show the total cost of running the inference function of BERT on different clouds. Deploying the workflow on $ali\_gpu$ results in maximum savings of 61.9\% and 82.2\% compared to $aws\_cpu$, depending on the batch size. Functions running on CPUs have longer execution times than those on GPUs; although CPUs may be cheaper, they ultimately cost more due to the increased runtime. Additionally, some serverless cloud providers do not offer GPU resources, and even when they do, the prices vary between clouds. This variability presents opportunities for optimization.



\subsection{Limitations of Existing Cross-cloud FaaS Workflows Orchestrators}
While FaaS model was initially used for simple applications with a single function, this paradigm has recently increasingly proved to be useful for complex applications consisting of many functions \cite{pu2019shuffling,jarachanthan2022astrea,wang2025edge}, i.e., combining them into function workflows. 
To organize complex function workflows in the expected order and ensure exactly-once semantics \cite{burckhardt2021durable,jia2021boki,zhang2020fault}, a common solution is to introduce a central orchestrator. Users need to logically define workflows and submit them to the orchestrator, e.g., by describing the interaction of each function. The orchestrator then realizes the defined workflow logic by invoking functions and managing runtime state (e.g., intermediate data and execution progress).

Recently, several studies \cite{ristov2022faast,ristov2021xafcl,khochare2023xfaas} have noted the performance limitations of single-cloud FaaS workflows and have explored using multiple CPU-based FaaS systems to optimize makespan. These cross-cloud orchestration solutions are almost all based on centralized designs (Master-Worker mode). The orchestrator acts as a coordination point for functions distributed across multiple FaaS systems. A centralized orchestrator can record the execution result of each function, decide which function to invoke next based on workflow progress, and avoid duplicate executions.
\begin{figure}[htbp]
\centering
\includegraphics[width=0.6\textwidth]{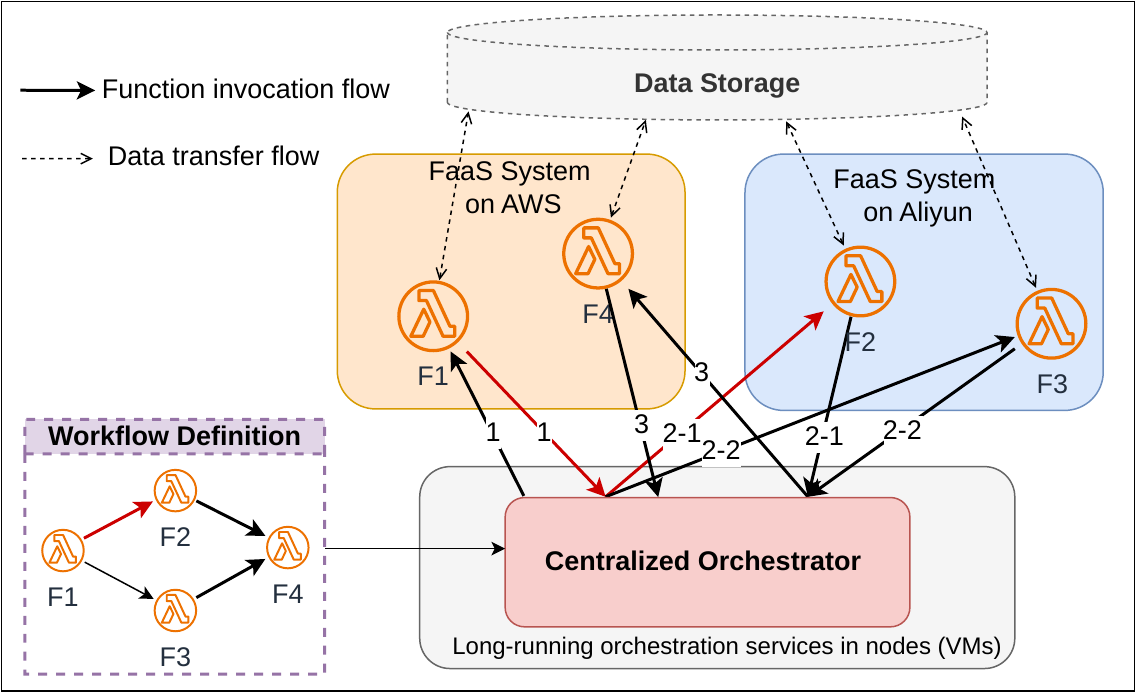} 
\caption{Logically centralized orchestrator organizing serverless workflows on Jointcloud FaaS systems.}
\label{centero} 
\end{figure}

However, centralized orchestrators have significant drawbacks when orchestrating serverless workflows on Jointcloud FaaS systems.
Functions in a workflow can be distributed across multiple clouds or regions, and the orchestrator is likely far away from most functions. Frequent cross-cloud state transitions therefore impose additional overhead on a centralized orchestrator. Figure \ref{centero} shows the extra overhead incurred when orchestrating cross-cloud workflows. Centralized state transitions add at least one additional communication link (red arrows in Figure \ref{centero}), resulting in non-negligible latency overhead and high data-egress charges. In addition, the logically centralized architecture limits the performance of highly concurrent tasks \cite{ristov2021xafcl} and incurs extra cost due to long-running orchestration services on nodes (e.g., virtual machines, VMs). External serverless workflow orchestrators \cite{awswf,gwf,azwf,aliwf} offered by CSPs also view workflows as state machines and charge separately (25\$ per 1M state transitions) for each state transition in a workflow, rather than for the actual resources consumed, which leads to unfair billing.

To address these limitations, we propose Joint$\lambda$, a distributed runtime system for cross-cloud heterogeneous serverless computing, accelerating makespan by leveraging inter-cloud heterogeneity, and function-side orchestration instead of centralized nodes.

\section{Related Work}
\textbf{Orchestrating serverless workflows on single-FaaS.} There is a long line of research \cite{fouladi2017encoding,ao2018sprocket,fouladi2019laptop,sreekanti20cloudburst,li2022faasflow,carver2020wukong,liu2023doing,burckhardt2022netherite} on serverless workflow orchestration for a single FaaS system. Some orchestrators \cite{fouladi2017encoding,ao2018sprocket,fouladi2019laptop,jonas2017occupy,sampe2018serverless} outsource work threads to cloud FaaS systems to accelerate tasks. The cloud function is designed as a generic worker. However, workers often suffer from large runtime initialization overhead. Some orchestrators \cite{sreekanti20cloudburst,azwf,aws,gcp,ali} are integrated with FaaS systems and enhance fault tolerance and scalability with message queues and databases. To guarantee the correct execution of the workflow, AFT \cite{zhang2020fault} proposes a fault-tolerant shim between the FaaS system and the datastore, while Olive \cite{setty2016realizing} and Boki \cite{jia2021boki} design fault-tolerant mechanisms based on function logs stored in the datastore. However, logically centralized controllers hinder orchestration performance. FaaSFlow \cite{li2022faasflow} presents a worker-side workflow scheduling pattern for serverless workflow execution, namely a distributed orchestrator at the VM level. Wukong \cite{carver2020wukong}, Pheromone \cite{yu2023following}, and Netherite \cite{burckhardt2022netherite} also optimize performance with decentralized orchestration. Unum \cite{liu2023doing} is an application-level,
decentralized orchestration system. Compared to master-worker orchestrators, it performs well, offers greater flexibility, and lowers expenses. Joint$\lambda$ integrates the above insights and introduces distributed orchestration and datastore-based execution guarantees for serverless workflows on Jointcloud FaaS systems.

\noindent \textbf{Orchestrator support for workflows on multiple FaaS.} \cite{malawski2020serverless} extends Hyperflow\cite{balis2016hyperflow} to run scientific computing serverless workflows on multiple clouds. Lithops \cite{lithops} views functions as homogeneous workers and leverages multi-cloud FaaS to optimize parallel tasks. To address the limitations of the number of concurrent functions, xAFCL\cite{ristov2021xafcl} implements a middleware that can schedule and execute different functions in a workflow across multiple FaaS. FaaSt \cite{ristov2022faast} added new features to xAFCL to enhance fault tolerance across multiple clouds. Globalflow\cite{zheng2019globalflow} uses connector functions to connect workflows across multiple regions in AWS. Based on the same idea, XFaaS \cite{khochare2023xfaas} presents a cross-platform orchestrator for FaaS workflows based on existing cloud orchestration services. To our knowledge, cross-cloud orchestrators for serverless workflows are logically centralized. However, Joint$\lambda$ is different from these systems.

\begin{table}[h!]
    \centering
    \caption{The design space of Joint$\lambda$ against existing works of serverless workflow orchestrator.}
    
    \begin{tabular}{c|c|c|c|c|c|c}
    \hline
     &  Joint$\lambda$ & xAFCL \cite{ristov2021xafcl} & Lithops \cite{lithops} & FaaSFlow \cite{li2022faasflow} & XFaaS \cite{khochare2023xfaas} & Unum \cite{liu2023doing} \\
    \hline
    \makecell{Supporting \\ Multi-FaaS?} & \Checkmark & \Checkmark & \Checkmark & \XSolidBrush & \Checkmark & \XSolidBrush \\
    \hline
    \makecell{Supporting \\ serverless workflow?}
     & \Checkmark & \Checkmark & \XSolidBrush & \Checkmark & \Checkmark & \Checkmark   \\
    \hline
    \makecell{Distributed \\or Centralized?} & Distributed & Centralized & Centralized & Distributed & \makecell{Semi-\\centralized} &  Distributed \\
    \hline
    
    \end{tabular}
     \label{compare_existing_works}
    \end{table}

    To summarize the above discussion, Table~\ref{compare_existing_works} compares Joint$\lambda$ with several representative systems along three dimensions. ``Supporting Multi-FaaS?'' indicates whether a system can operate across multiple FaaS platforms. ``Supporting serverless workflow?'' indicates whether it natively supports workflow-oriented control flow, such as sequences, DAGs, fan-out/fan-in, or other explicit dependencies among functions, rather than treating functions only as independent or embarrassingly parallel tasks. ``Distributed or Centralized?'' indicates whether the orchestration architecture is distributed or relies on a centralized coordinator.

    Beyond the specific domain of workflow orchestration, recent literature highlights the rapid evolution of the broader FaaS ecosystem. For instance, to facilitate serverless research, tools like faas-sim \cite{raith2023faas} have been developed for platform simulation, whereas Joint$\lambda$ focuses on building a real-world runtime system. From an architectural perspective, Zhu et al. \cite{zhu2024radf} proposed RADF to modularize FaaS platforms for better maintainability. Furthermore, the application landscape of FaaS is expanding into complex workloads, such as distributed deep neural network inference across the cloud-to-things continuum \cite{bueno2024functions}. These structural and applicative advancements demonstrate the growing demand for flexible, heterogeneous FaaS environments, a landscape that Joint$\lambda$ further enriches by providing robust, cross-cloud workflow orchestration with exactly-once execution semantics.

\section{RATIONALE of Joint$\lambda$}

This section introduces Joint$\lambda$, a function-side workflow orchestration system designed for orchestrating serverless workflows on Jointcloud FaaS systems. Compared to existing systems, Joint$\lambda$ aims to achieve the following underexplored goals: 1) Joint$\lambda$ leverages inter-cloud heterogeneous FaaS systems, such as those supporting GPU accelerators, to accelerate completion time and reduce costs. 2) Instead of relying on centralized orchestration nodes to manage and execute workflows, Joint$\lambda$ uses function-side orchestration to enable independent function invocations and data transfers, reducing the overhead of cross-cloud communication.
3) Joint$\lambda$ supports exactly-once execution semantics for workflows, introducing fault tolerance and failover in unstable multi-FaaS environments.
\subsection{Architecture Overview}

\begin{figure}[htbp]
  \centering
  \includegraphics[width=0.8\textwidth]{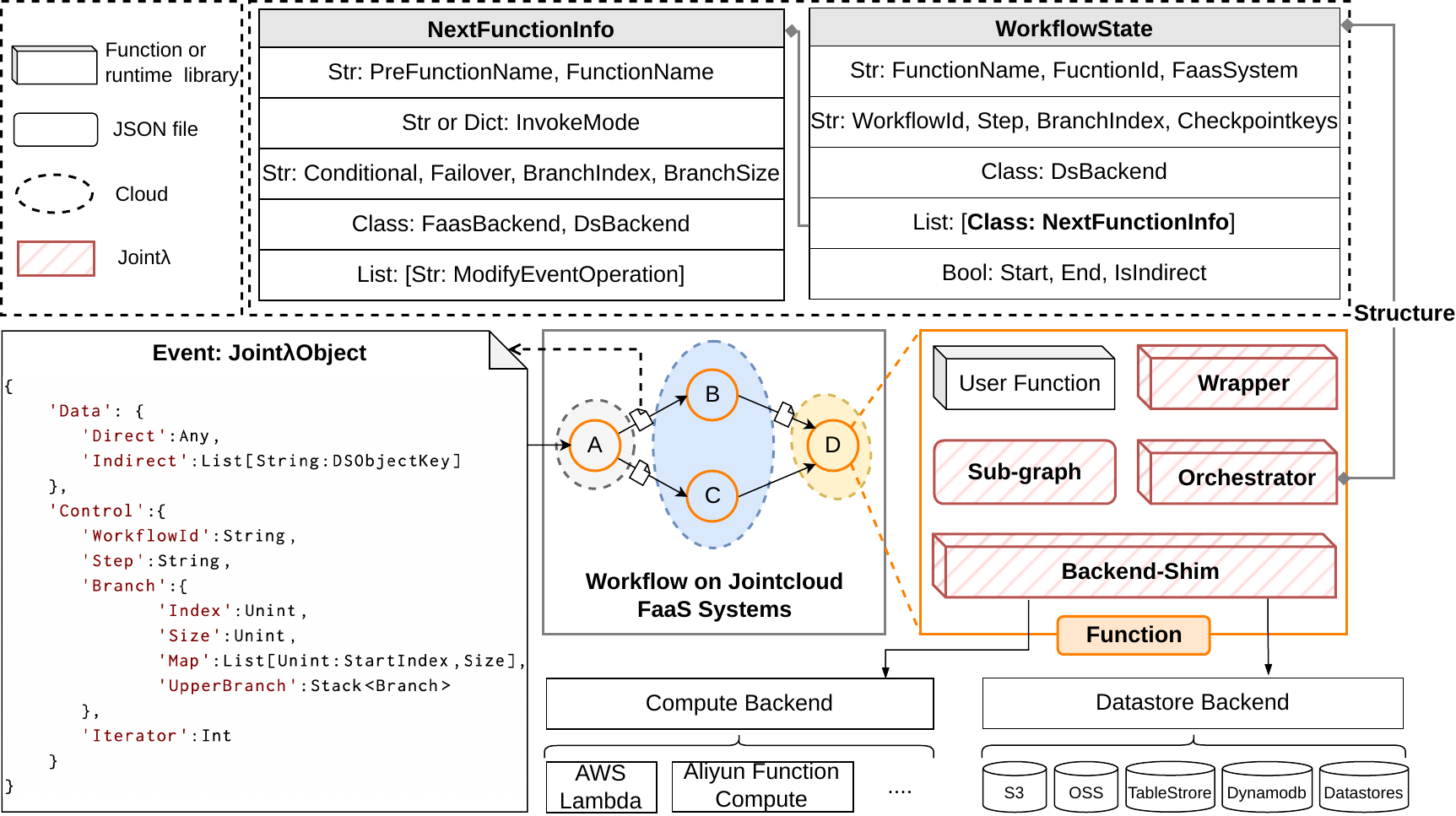} 
  \caption{Abstraction for Joint$\lambda$. The data structure for the management of function invocation and data transfer in the function-side workflow orchestrator.}
  \label{abs} 
  \end{figure}
Figure \ref{abs} provides an architectural overview of the three main components of Joint$\lambda$: the wrapper, the function-side orchestration library, and Backend-Shim. The developer writes a sub-graph for each function based on the workflow logic. The wrapper handles function entry and exit, parses the incoming $Joint\lambda Object$, passes user data to the function, and encapsulates the output for downstream calls. The orchestration library then executes the invocation and data-transfer logic specified by the sub-graph, while Backend-Shim maps these operations to the native APIs of the target FaaS and storage backends.
\subsection{Backend-Shim}


As shown in Figure \ref{abs}, Backend-Shim is a compatibility layer designed as an independent library that allows each function to invoke downstream functions and access storage through a uniform interface. It currently supports functions built on the Python runtime and is composed of Compute-Backend-Shim and DataStore-Backend-Shim. 
 First, Backend-Shim is dedicated to shielding cloud heterogeneity, providing a unified abstraction for function execution and data access. The function-side workflow orchestrator accesses various data storages and cross-cloud function invocations through Backend-Shim. 
 Second, Backend-Shim is implemented based on the APIs of different FaaS systems and data storages, providing a standard interface for the function-side workflow orchestrator, as shown in Table \ref{shim}.

\begin{table}[!htbp]
\caption{Backend-Shim APIs. Object storage and table storage in DSBackend implement the red APIs for indirect data transfer.}\label{shim}
\centering

\begin{tabular}{@{}lll@{}}
\toprule
\textbf{Class}                         & \textbf{API}                                    & \textbf{Description}                                   \\ \midrule
                              & { create(ds)}       & Create a datastore client.                    \\
 &
  {store\_output\_data(key, data)} &
  \begin{tabular}[c]{@{}l@{}}Conditionally create an item/object, \\ i.e. create an item if it does not exist.\end{tabular} \\
                              & { get\_value(key)}  & Strong consistency read an item/object.       \\
 &
  create\_invocation\_list(key) &
  \begin{tabular}[c]{@{}l@{}}Conditionally create a list of strings, \\ i.e. create an item if it does not exist.\end{tabular} \\
                              & append\_and\_get\_list(key, func\_list) & Append list items and return the latest list. \\
 &
  create\_bitmap(size, key) &
  \begin{tabular}[c]{@{}l@{}}Conditionally create a bitmap, \\ i.e. create an item if it does not exist.\end{tabular} \\
\multirow{-9}{*}{\textbf{DSBackend}}   & update\_bitmap(index, key)              & Update a bitmap by corresponding position.    \\ \midrule
                              & create(faas)                            & Create a FaaS client.                         \\
\multirow{-2}{*}{\textbf{FaaSBackend}} & async\_invoke(function, payload)        & Asynchronous http invocation.                 \\ \bottomrule
\end{tabular}
\end{table}

  We abstract data operations into a set of unified APIs as shown in Table~\ref{shim}. These APIs are categorized into two groups based on their roles in the workflow: 1) Data Transfer: Primitives such as \texttt{store\_output\_data} and \texttt{get\_value} are used for large payload persistence.  In practice, object storage mainly carries large intermediate payloads, whereas table storage maintains checkpoint metadata and runtime coordination state. 2) Workflow Coordination: To support complex patterns like fan-in, we implement specialized primitives including atomic bitmaps and invocation lists. For instance, \texttt{update\_bitmap} is invoked by parallel upstream functions to signal completion, enabling the shim layer to trigger downstream tasks without a centralized scheduler, whose detailed design is described in Section~\ref{sec:collaboration}.


Our current prototype implements Backend-Shim as a Python library embedded in the Joint$\lambda$ wrapper, leveraging mature Python runtimes and stable SDKs on mainstream FaaS platforms such as AWS Lambda and Aliyun FC. Porting Joint$\lambda$ to another language mainly requires equivalent wrapper/shim bindings while preserving the same sub-graph IR and Joint$\lambda$Object schema, including (i) Joint$\lambda$Object (de)serialization, (ii) the small set of \texttt{DSBackend} atomic primitives, and (iii) an adapter to the platform's asynchronous invocation API; BYOR/container-image packaging (e.g., AWS Lambda container images) can bundle the runtime and dependencies, although practical portability is still constrained by provider differences in event models and SDK feature sets.

\subsection{Distributed Orchestration Capabilities}

In Joint$\lambda$, there is no centralized orchestration node managing the workflow state and executing workflows (data transfer, function invocation). Each function obtains a local \textit{sub-graph} of the workflow and the runtime state of the workflow (\textit{WorkflowState}) through a function-side workflow orchestrator to perform function invocations, data transfer, and other orchestration logic, as shown in Figure \ref{abs}. Since the sub-graph may contain complex structures and functions may be distributed across multiple FaaS systems, designing and organizing data structure is crucial for orchestrating workflows.

\textbf{Structure organization in function-side workflow orchestrator.} In the workflow, functions need to maintain their own workflow state and invoke functions across different FaaS systems while transferring data upon completion. To allow each function to independently manage orchestration logic, the \textit{WorkflowState} structure was introduced, as shown in Figure \ref{abs}. This structure acts as the maintainer of the workflow sub-graph and runtime state. The \textit{WorkflowState} saves the workflow runtime state obtained from input events, including unique \textit{WorkflowId}, execution stage \textit{step}, and branch number \textit{branchIndex}. Additionally, the \textit{WorkflowState} structure records the metadata of the current and all subsequent functions within the local sub-graph, where \textit{NextFunctionInfo} maintains the metadata for each subsequent function.

\textbf{Data and Control Transfer with $Joint \lambda Object$.} To support the transfer of workflow runtime data and control information across multiple FaaS systems, a new cloud event object, $Joint \lambda Object$, and a wrapper were introduced. The format of $Joint \lambda Object$ is depicted in Figure \ref{abs}. Within the wrapper, the Unwrap function retrieves input data from the incoming $Joint \lambda Object$ and passes it to the user function, while generating a unique ID for the current function based on the 'Control' field in $Joint \lambda Object$. The input data is either transferred directly via the Direct field of $Joint \lambda Object$ or pulled indirectly by the Unwrap function from data storage, which are also output checkpoints of previous functions. The unique ID is primarily used for naming storage items as checkpoints and for cross-cloud collaboration points. The Wrap function generates a new $Joint \lambda Object$ for each subsequent function invocation based on the current workflow state.


\textbf{The design of sub-graph, invocation primitives and transfer primitives.} FaaS workflows are commonly represented as DAGs, where functions are defined as work nodes and edges represent data flows between these functions. Widely used FaaS orchestrators, such as AWS Step Functions, often utilize high-level descriptive languages to express interactions between function nodes. This concise workflow invocation primitive is sufficient to represent basic patterns such as direct invocation (Sequence, Parallel), dynamic invocation (Map), and conditional invocation (Cycle, Fan-In, Choice). Consequently, we continue to use this form to express serverless workflows across multiple FaaS systems.

\begin{figure}[htbp]
\centering
\includegraphics[width=0.6\textwidth]{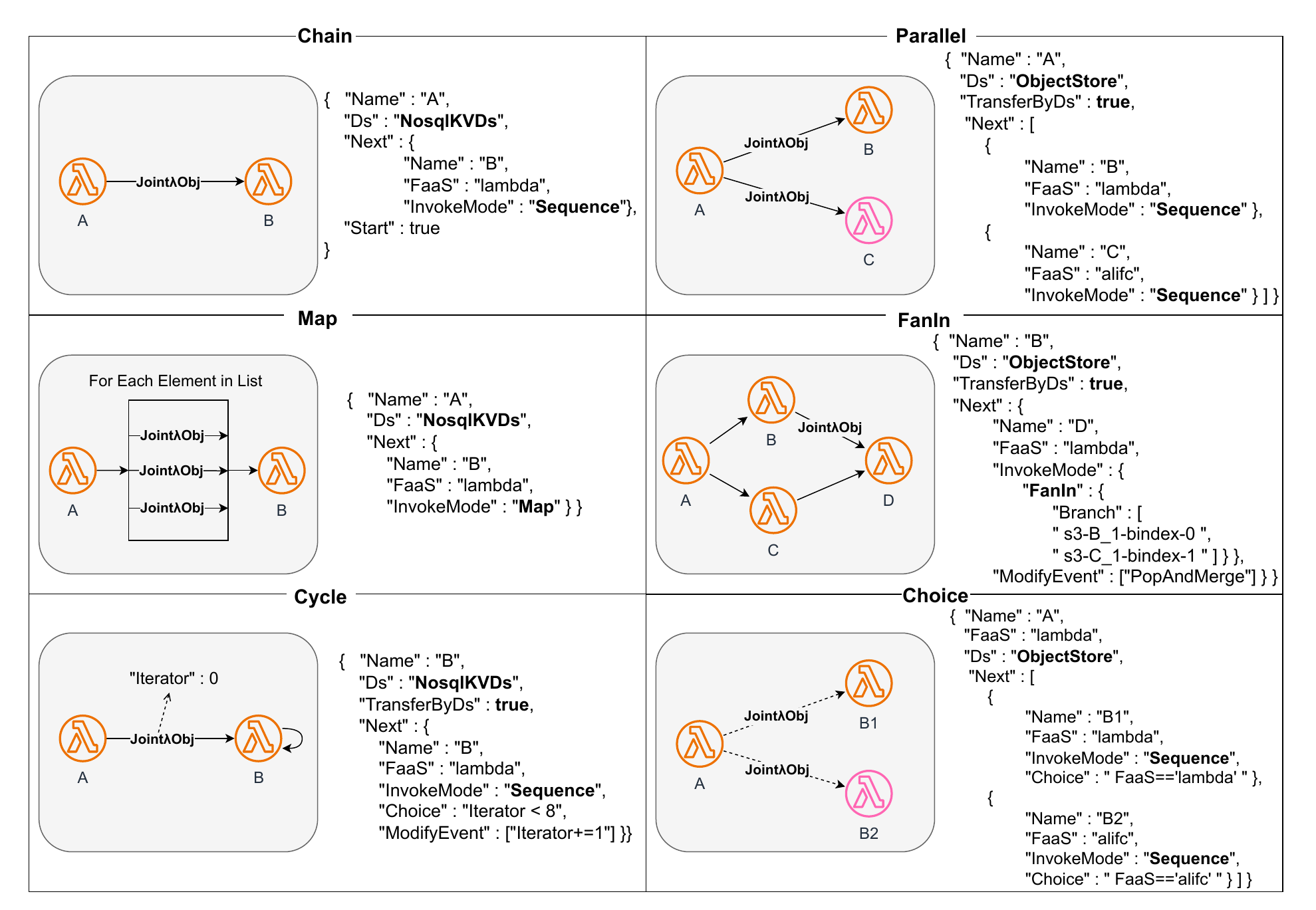} 
\caption{Sub-graph definitions and major primitives for workflow basic patterns. Invocation primitives and data transfer primitives (bold text in the figure) can be combined flexibly.}
\label{basicpattern} 
\end{figure}

Due to the absence of a global graph in function-side workflow orchestration, a local sub-graph containing invocation primitives and data transfer primitives is introduced to provide metadata for functions and data storage within the sub-graph. Figure \ref{basicpattern} illustrates the sub-graph definition of basic workflow patterns. The function-side workflow orchestrator creates \textit{NextFunctionInfo} objects based on the subsequent function's FaaS system and invocation primitives, and creates checkpoints for fault tolerance and data storage backends for indirect data transfer through data transfer primitives.

Joint$\lambda$ implements fault tolerance via idempotent checkpoints backed by provider storage primitives: object storage (e.g., S3 \cite{aws_s3}/OSS \cite{alibabacloud_oss}) for large intermediate objects and table/kv storage for checkpoint metadata and coordination. For payload handling, Joint$\lambda$ uses direct transfer for small payloads and indirect transfer through object storage with a stable reference carried in the Joint$\lambda$Object for large payloads or when request sizes exceed provider limits.

Invocation primitives can be flexibly combined with data transfer primitives. Among these combinations, patterns that coordinate multiple inputs, such as fan-in, typically require indirect transfer. To reduce cross-cloud data egress in such cases, Joint$\lambda$ selects the data placement backend at runtime, typically choosing the cloud where the majority of functions in the sub-graph are located.

\begin{figure}[htbp]
\centering
\includegraphics[width=0.6\textwidth]{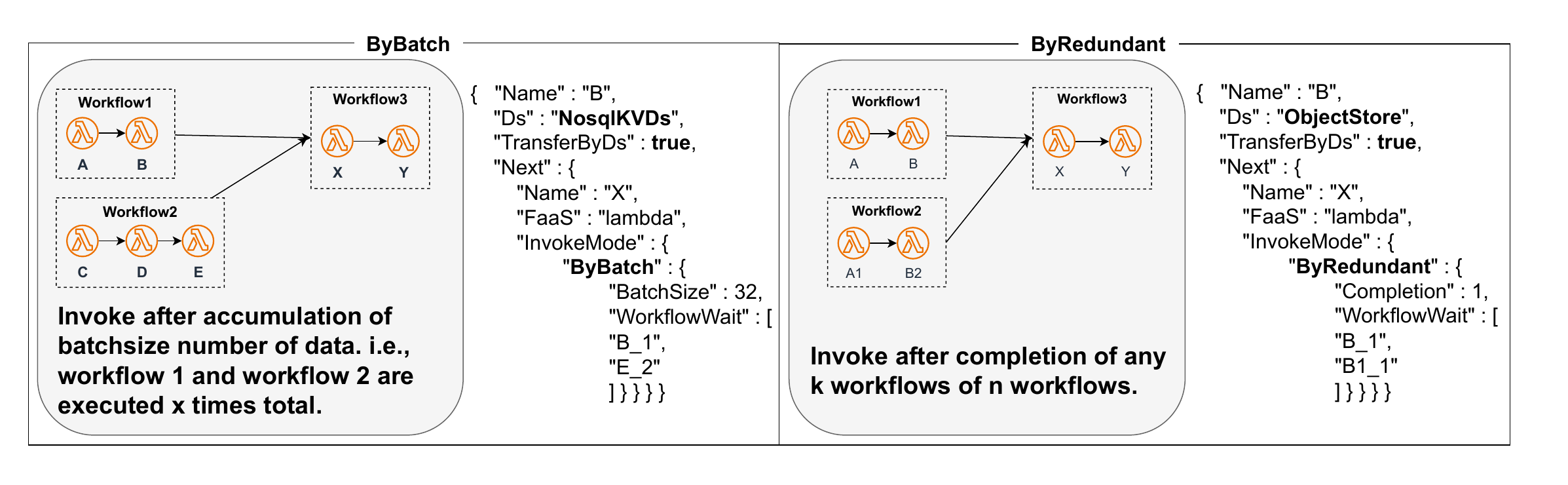} 
\caption{Sub-graph definitions and major primitives for workflow collaboration across time and space.}
\label{workflowcollaboration} 
\end{figure}

Additionally, to introduce greater flexibility, we have also designed invocation primitives to enable multiple workflows in different spaces to collaborate across different temporal dimensions, as shown in Figure \ref{workflowcollaboration}. The batch invocation primitive invokes other workflows to process after accumulating a specific amount of data across multiple workflows. The redundancy invocation primitive is utilized to manage redundant requests to mitigate the effects of stragglers. A key insight is that geographically distributed functions can achieve local collaboration through strongly consistent data storage. However, this increased flexibility requires users to meticulously design the interactions between workflows.


\section{Design of Joint$\lambda$}


In this section, we introduce how Joint$\lambda$ achieves two important objectives in serverless workflow orchestration across multiple FaaS systems: exactly-once execution semantics and distributed collaboration.

\subsection{Exactly-once Execution}
In a distributed computing environment that combines multiple FaaS systems, serverless workflow functions run across different cloud regions and FaaS systems. Intermediate data is distributed and transferred across multiple cloud storage systems. As a result, orchestrating serverless workflows on Jointcloud FaaS systems faces increased instability, such as cross-cloud communication link failures or computing service outages. Functions can fail at any execution stage, but FaaS systems typically offer weak execution guarantees. Most FaaS systems only provide \textbf{at-least-once} execution guarantees, meaning that if a failure occurs, retries might result in the same function executing multiple times. Specifically, when a FaaS system receives an asynchronous invocation request, it doesn't execute the function immediately but responds to the invoker and queues the request, triggering the function via a message queue. If the function crashes, the FaaS system can retry the execution using the persisted request. In Joint$\lambda$, the execution semantics are ensured by a checkpointing protocol based on conditional writes and strongly consistent reads. Optional storage features such as S3 Object Lock may provide additional hardening, but they are not required by our prototype.
To ensure exactly-once execution semantics in such a challenging environment, Joint$\lambda$ treats function execution as two stages: data production and function invocation. It saves checkpoints using APIs outlined in Table \ref{shim} to guarantee that each stage executes only once. Given the at-least-once execution guarantees of the FaaS systems, together with at-most-once data production and at-most-once function invocation, we achieve exactly-once execution semantics. Next, we explain how we achieve at-most-once data production and at-most-once function invocation.

\subsubsection{Output Data Checkpoints}
Joint$\lambda$ calculates the globally unique ID (\textit{FunctionId}) of the function based on the runtime state of the $Joint \lambda Object$ and adds a suffix to derive the key for the output data checkpoint (\textit{outputKey}). It attempts to retrieve the output result checkpoint of the function. If the checkpoint exists, the checkpoint data is used as the output. Otherwise, it executes the user function and uses $store\_output\_data$ within the $Wrap$ function to save the output data checkpoint. To accommodate user code containing logic of uploading data, Joint$\lambda$ exposes the outputKey to the user function and uses the \textit{isStored} flag to indicate whether it has been stored. The checkpoint ensures that the output data from the first successful execution is transferred, and repeated executions do not alter the execution result, achieving at-most-once data production. We currently do not rely on instance-local \texttt{/tmp} cache as a correctness-critical optimization. Since \texttt{/tmp} is non-durable and local to a specific warm instance, retries, failover, or cross-cloud re-execution may observe a different instance where the cache is absent or stale. We therefore treat storage-backed checkpoints as the source of truth; \texttt{/tmp} can still be used as a best-effort cache if validated against checkpoint metadata such as the output key or object Entity Tag (ETag).

\subsubsection{Invocation Checkpoints.}
 Similarly, Joint$\lambda$ appends a suffix to the globally unique ID (\textit{FunctionId}) of the function to derive the key (\textit{ivkKey}) for the invocation checkpoint. The invocation checkpoint is designed as an initially empty string list stored in a table storage system on the cloud where the function resides. Joint$\lambda$ first queries whether the invoked function has been recorded in the checkpoint. If it doesn't exist, it modifies the $Joint \lambda Object$ and runs the invocation logic based on the \textit{InvokeMode} of the invoked function. Joint$\lambda$ uses asynchronous invocation (\texttt{FaaSBackend.async\_invoke}) for cross-platform downstream calls to avoid double-billing (charging the invoker while it waits and the callee while it runs). Upon successful invocation of the subsequent function, it appends the name of the successfully invoked function(s) to the invocation checkpoint. The checkpoint ensures recording the first successful invocation of the function, achieving at-most-once function invocation.
In practical running, we encountered unacceptable delays when invoking a large number of concurrent functions. Upon investigation, we identified that the performance degradation stemmed from extensive checkpoint reads and writes, as well as asynchronous invocation delays. Therefore, Joint$\lambda$ optimizes the latency of reading and writing to table storage for all concurrent invocations (fan-out) exceeding 10 subsequent functions by \textbf{grouping checkpoint}. Joint$\lambda$ also utilizes concurrent invocations with 10 threads for fan-out. It invokes every 10 functions in the order of $nextFuncs$, and upon successful invocation, appends them to the invocation checkpoint as a group.


In the most extreme scenario, a function crashes and retries after executing an asynchronous invocation statement, without
yet saving the invocation checkpoint for that function, which leads to duplicate invocations. However, Joint$\lambda$ ensures that the
function uses the same data through output result checkpoints and guarantees at-most-once invocation by means of repeated
function invocation checkpoints. Thus, even in extreme cases of duplicate function invocations, it does not affect the exactly-once
execution of subsequent workflow steps, ensuring correct execution.


\subsection{Failover}
\begin{figure}[htbp]
\centering
\includegraphics[width=0.45\textwidth]{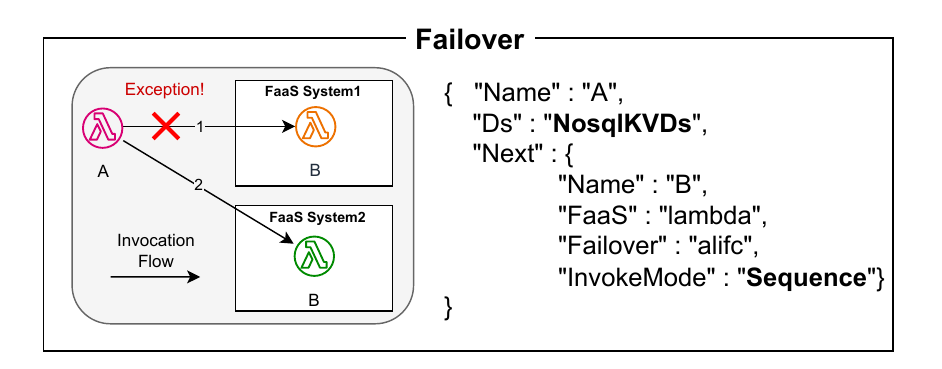} 
\caption{An example of failover and its sub-graph definition.}
\label{fo} 
\end{figure}

Failover leveraging multi-cloud environments are viable solutions for enhancing availability. However, they often rely on manual backup procedures and operations personnel for recovery in traditional cloud computing paradigms, facing limitations such as low resource utilization and high costs. Fortunately, the serverless paradigm's on-demand billing and automatic scaling eliminate these drawbacks. Backup functions in workflows consume no resources and incur no costs until they are invoked. Leveraging these insights, Joint$\lambda$ implements failover during the function invocation stage, efficiently and automatically transferring workflows in case of FaaS system failures, as depicted in Figure \ref{fo}. The "Failover" field of functions in the sub-graph specifies alternative FaaS systems, indicating pre-deployed backups (e.g., B of FaaS System2 in Figure \ref{fo}).

 When a function fails to invoke subsequent functions (possibly due to FaaS system failures or network issues), Joint$\lambda$ does not immediately fail but rather captures the exception and enters the exception handling process. Joint$\lambda$ then uses the Backend-Shim to create a client for a backup FaaS system and re-invokes the function on the backup FaaS platform via asynchronous invocation (\texttt{async\_invoke}). This capability allows Joint$\lambda$ to seamlessly migrate workflows in multi-cloud environments, enhancing availability without compromising cost or resource utilization.

\subsection{Data Transfer and Collaboration}

\subsubsection{Cross-platform Data Transfer}



The "TransferByDs" primitive in the sub-graph determines the transfer method, while the "Ds" primitive specifies the type of data storage in Figure \ref{basicpattern}. Joint$\lambda$ supports \textbf{direct data transfer} through the $Joint \lambda Object$ in HTTP requests. For direct transfer, Joint$\lambda$ creates output data checkpoints on storage services co-located with the current FaaS platform, e.g., using AWS Lambda together with Amazon S3 or Amazon DynamoDB.
\begin{figure}[htbp]
\centering
\includegraphics[width=0.48\textwidth]{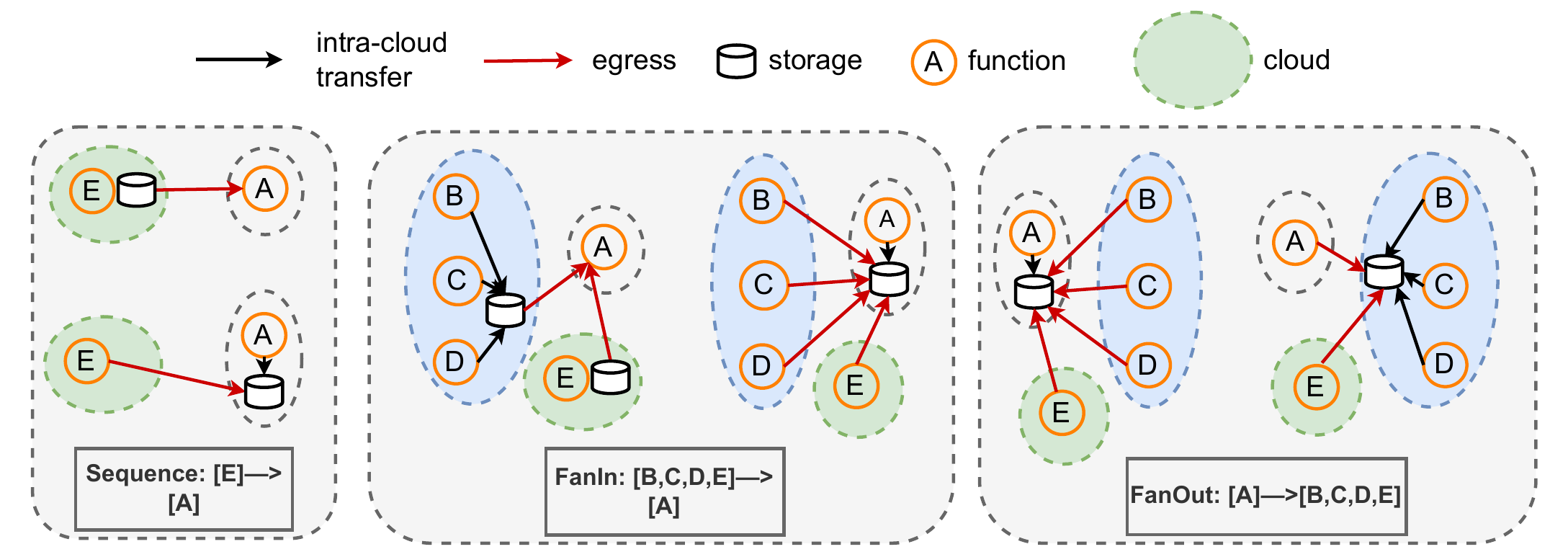} \label{transfer_seq} 
\caption{The indirect data transfer for three invocation modes with different data placement.}\label{transfer}
\end{figure}
Joint$\lambda$ also supports \textbf{indirect data transfer} via external storage when the request payload exceeds the hard quota of the FaaS platform (e.g., the size limit for asynchronous request payloads is 256 KB on AWS Lambda and 128 KB on Aliyun Function Compute). For indirect transfer, we aim to reduce data egress fees through strategic data placement. Analysis of data flows reveals that 1:1 (sequence) and n:1 (fan-in) invocation relationships cannot reduce the amount of egress traffic through data placement, as shown on the left and in the middle of  
 Figure \ref{transfer}. However, concurrent 1:n invocations (fan-out) can optimize the number of egress data through data placement. 
We have devised a simple yet effective method for selecting the location of data storage based on the \textbf{majority rule principle}. Specifically, Joint$\lambda$ first counts the number of FaaS systems in the sub-graph and then selects the most frequent FaaS system to create data storage at the same location. This data storage is used to store the output data checkpoints, from which the next functions can read the data. As shown in the right of Figure \ref{transfer}, we can reduce the number of egress data by placing the data in the cloud where B, C and D functions are located.
\subsubsection{Cross-platform Function Collaboration}
\label{sec:collaboration}

\begin{figure}[htbp]
\centering
\includegraphics[width=0.45\textwidth]{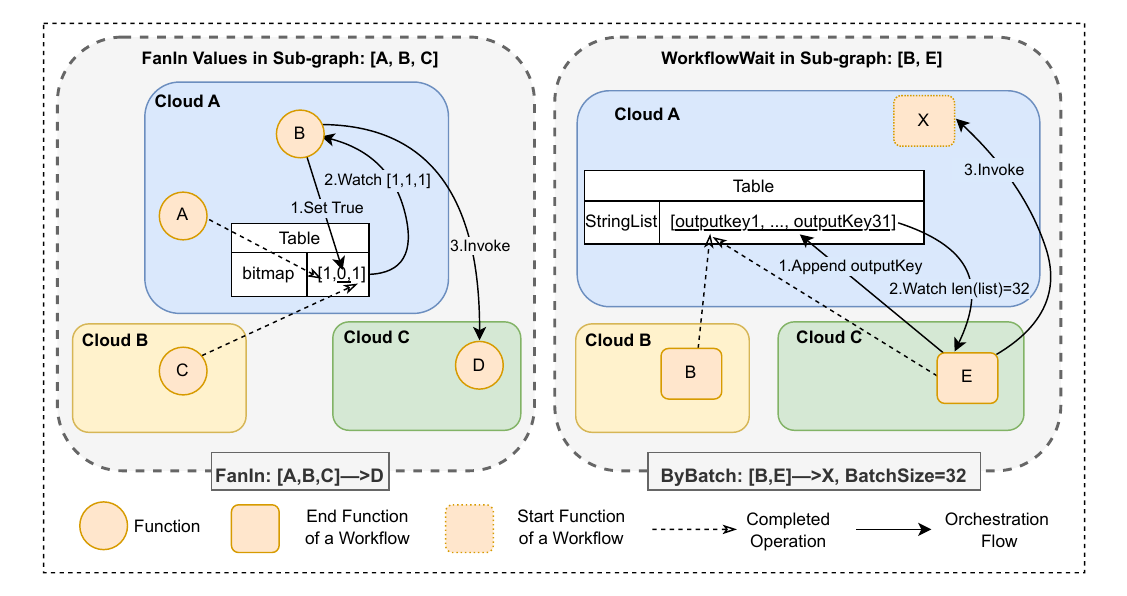} 
\caption{Cross-platform functions collaborate through partial coordination point.}
\label{colla} 
\end{figure}

In distributed serverless Workflows on Jointcloud FaaS Systems, functions cannot directly communicate to check each other's completion status. Therefore, multiple functions working together (e.g., invoking an aggregation function after accumulating four invocation requests) require a local coordination point. We believe that the \textit{List} data type in strongly consistent NoSQL \textit{Key-Value} storage is sufficient to serve as the local coordination point. Additionally, the sub-graph lists the metadata of all collaborating functions in the invocation primitives. Figure \ref{colla} shows how Joint$\lambda$ organizes the orchestration process for collaboration.

Using the fan-in within the workflow on the left figure as an example, we explain how Joint$\lambda$ uses the bitmap primitive in Table~\ref{shim} as the coordination mechanism. Joint$\lambda$ appends a suffix (e.g., -bitmap) to the unique ID of the aggregation function (function D in the figure) to obtain the coordination point's key and creates the Boolean bitmap as an item in the DSBackend table-storage (i.e., AWS DynamoDB \cite{awsdynamodb} or Aliyun TableStore \cite{aliyuntablestore}), and is strategically co-located with the cloud provider that hosts the majority of the fan-in branches to minimize cross-cloud communication overhead. The boolean bitmap indicates whether a function has been executed. Functions A, B, and C update their corresponding boolean values in the bitmap to True after execution and perform a strongly consistent read of the bitmap. The function (B) that reads the target value (all elements of the bitmap being True) will invoke the aggregation function (D).


\textit{ByBatch} and \textit{ByRedundant} on the right figure follow a similar orchestration process. However, to accommodate dynamics, the coordination point is set as a string List and is created in the same data storage as the invoked function. The key of the coordination point differs fan-in within the workflow on the left figure, using the concatenated names of all functions in the sub-graph as the key.

\subsection{Unique Key and Function's Reusability} 

\begin{figure}[!htbp]
\centering
\includegraphics[width=0.34\textwidth]{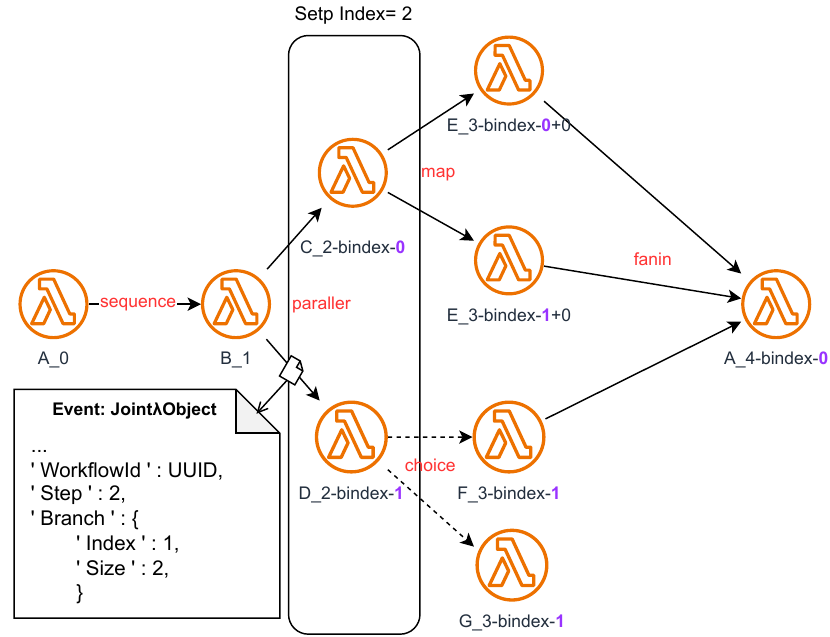} 
\caption{Unique Id of functions in workflow. The \textit{workflowId/} prefix is omitted.}
\label{id} 
\end{figure}

Each workflow generates a workflow ID using a UUID in the initial function, and this ID is passed through the $Joint \lambda Object$. The workflow ID serves as a common prefix for all names in the current workflow invocations, facilitating garbage collection after completion. Although the use of global function names and workflow IDs can distinguish different functions in different workflows, this is not sufficient for unique function naming. A single workflow may invoke the same function multiple times, such as in mappings, loops, or repeated usage. Joint$\lambda$ introduces step index and branch index to uniquely name function nodes. The step index increments with each step, while the branch index depends on the Branch field in the $Joint \lambda Object$, relating to fan-out and fan-in operations. By calculating unique function names, Joint$\lambda$ supports unique data storage keys, ensuring the correctness of orchestration and the reusability of functions.
As illustrated in Figure \ref{id}, Joint$\lambda$ appends a step index to function names at the same stage. Joint$\lambda$ uses a branch index to distinguish fan-out branches. For example, functions C and D are at step 2, located in branches 0 and 1, respectively, thus named C\_2-bindex-0 and D\_2-bindex-1. Joint$\lambda$ uses "+" to distinguish multiple branch levels. During a fan-out invocation, Joint$\lambda$ adds the corresponding Branch stack to the function's name, as seen with function B. The purple numbers in the figure represent the new numbers to stack, with 0 and 1 are pushed onto name stacks of functions C and D, respectively. The Branch field continues to propagate until encountering a fan-in.
The fan-in invocation primitive is used with \textit{PopAndMerge}, as fan-in involves merging at least one branch level, modifying the Branch field. For example, with functions F and E, Joint$\lambda$ names functions according to the order of the fan-in Branch in the sub-graph: E\_3-bindex-0+0, E\_3-bindex-1+0, and F\_3-bindex-1. Then, it calculates the unique ID for function A. First, Joint$\lambda$ pops one level from all fan-in branches (e.g., from bindex-1+0 to bindex-0), resulting in E\_3-bindex-0, E\_3-bindex-0, and F\_3. Next, Joint$\lambda$ merges the fan-in branches. The merge selects the highest index and level branch index and updates the Branch stack in the $Joint \lambda Object$. Thus, this branch index becomes part of the unique ID for function A, i.e., A\_4-index-0. Joint$\lambda$ treats Choice invocation the same as Sequence invocation within a single branch.

 \textbf{Garbage collection in a workflow.}
During orchestration, Joint$\lambda$ stores data across multiple data storage systems. The intermediate data generated by FaaS workflows has a very short lifecycle because it is typically only valid within the current workflow. If left unmanaged, this temporary data will accumulate with each workflow invocation, leading to increased storage costs. The distribution of intermediate data across multiple clouds presents additional challenges for garbage collection. To address this, Joint$\lambda$ introduces a garbage collection mechanism.  The GC function is an independent, parallel function that can automatically scale up or down based on demand.

Joint$\lambda$ deploys Garbage Collection (GC) functions on compatible FaaS systems. The garbage collector is designed to be invoked by the workflow's terminal function. Once the workflow ends, all GC functions in the cloud are invoked concurrently. The GC functions extract the workflow ID from the received $Joint \lambda Object$ and heterogeneously clean all data storage items with the workflow ID as their prefix, based on different storage backends. This process effectively cleans data across various cloud storage systems since all data items produced by a workflow share a common prefix.

While object storage services such as Amazon S3 provide lifecycle rules for automatic expiration, Joint$\lambda$ retains an explicit GC mechanism for two reasons. First, lifecycle rules are typically coarse-grained (e.g., one day), whereas our GC reclaims short-lived workflow data immediately after completion. Second, Joint$\lambda$ spans multiple clouds and storage backends, so an independent GC function provides a unified cleanup mechanism across both object stores and table storage.

\section{Evaluation}
\subsection{Experimental Setup}



\noindent In this section, our evaluation answers the following questions: \vspace{-0.3em}

\begin{itemize}
 \item \emph{Question-1}: Can Joint$\lambda$ achieve better performance and lower cost than serverless workflows on a single FaaS system? (§\ref{section_benefits})
  \item \emph{Question-2}: Can Joint$\lambda$ enable efficient failover? (§\ref{section_recovery})
  \item \emph{Question-3}: How does Joint$\lambda$ compare with state-of-the-art cross-cloud serverless workflow orchestrators? (§\ref{section_Optimization})
  \item \emph{Question-4}: What latency overheads are incurred when orchestrating workflows with Joint$\lambda$? (§\ref{section_Overhead})
\end{itemize}
\noindent \textbf{Setup.} We implement Joint$\lambda$ on two FaaS platforms, AWS Lambda and Aliyun Function Compute (FC), including configurations with GPU accelerators. We use fully managed NoSQL storage services provided by the CSPs to store checkpoints and coordination points, i.e., DynamoDB \cite{awsdynamodb} and TableStore \cite{aliyuntablestore}. In our experiments, we do not modify the user functions' upload and download logic; therefore, large data objects such as videos and datasets are still transferred through object storage (Amazon S3 and OSS, i.e., Alibaba Cloud Object Storage Service). The output data checkpoints record only the intermediate data produced by user functions. All services are co-located in the same region (ap-northeast-1 for AWS and ap-north-1 for Aliyun), except for Aliyun CloudFlow, which is deployed in us-west-1 due to regional support limitations. Each workflow uses a separate table and is warmed up in advance to reduce the effect of cold starts.
 
\noindent \textbf{Baseline.} We compare  Joint$\lambda$ with the following baselines.
\begin{itemize}
  \item \textbf{AWS Step Functions (ASF) \cite{awswf}:} We use the ASF standard pattern to orchestrate serverless workflows on AWS Lambda as it implements exactly-once execution semantics.

  \item \textbf{Aliyun CloudFlow (AC) \cite{aliwf}:} AC also uses the state-machine model and implements exactly-once execution semantics. It is used to orchestrate serverless workflows on Aliyun FC. 

  \item \textbf{xAFCL \cite{ristov2021xafcl,ristov2022faast}:} xAFCL implements a middleware that can
schedule and execute different functions in a workflow across multiple FaaS systems, with centralized orchestration. It consists of an orchestrator and a database node.

\item \textbf{XFaaS \cite{khochare2023xfaas}:} XFaaS is a cross-platform orchestrator for serverless workflows based on connector-based cloud orchestration services. However, because the connector is a message queue, it only supports linear compositions. XFaaS is only used to evaluate sequence workflows. 
\item \textbf{Lithops \cite{lithops,sanchez2020primula,sampe2018serverless}:}  Lithops is a distributed computing framework that leverages multi-cloud FaaS systems. It allows local embarrassingly parallel tasks to be accelerated by outsourcing. It views functions as homogeneous workers, i.e., cloud threads. Workers pull code and data before running. Lithops is only used to evaluate parallel tasks in workflows.
\end{itemize} 
\begin{figure}[htbp]
\centering
\includegraphics[width=0.5\textwidth]{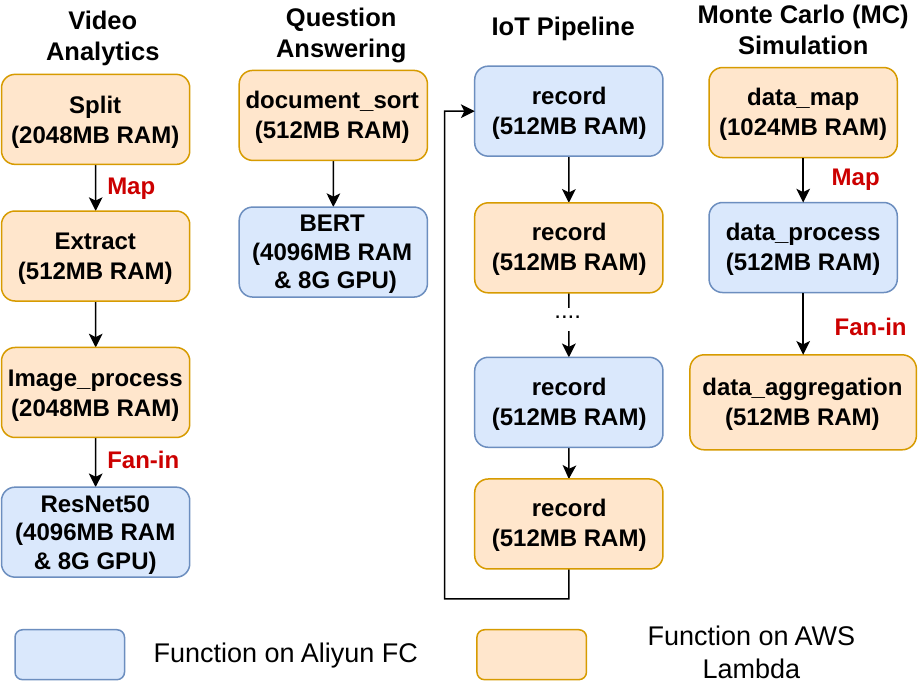}
     \caption{DAGs of the serverless workflows in the evaluation.}  \label{workload} 
\end{figure} 
\noindent \textbf{Workload.} Our experiments use four representative serverless workflows. These workflows have
different characteristics in terms of structure, data
dependencies, and FaaS systems. Figure \ref{workload} shows the DAGs of workflows, and the details are as follows.
\begin{itemize}

\item \textbf{Video Analytics} is based on our modified Orion\cite{mahgoub2022orion}, which consists of four stages: video split, frame extraction, image processing, and image recognition. Each workflow invocation processes a 1-minute video with 4 or 8 parallel branches.

\item \textbf{Question Answering Inference} is a batch task with the raw document  datasets stored in Amazon S3. It consists of two stages: article and question sorting, and QA inference\cite{serverlessbert}. Each workflow execution infers 4 QA and transfers about 40KB data.

\item \textbf{IoT Pipeline} is a synthetic sequence workflow in which each stage simply receives a small IoT payload, returns it unchanged, and lets the runtime checkpoint and asynchronously forward it to the next function. We vary the number of stages in the chain, and the intermediate payload size is fixed at 1 KB. 

\item \textbf{Monte Carlo (MC) Simulation} is adopted from xAFCL\cite{ristov2021xafcl} and is often used for numerical simulations in scientific computing. Each workflow execution generates 1 million numbers (data\_map). We estimate the PI value (data\_aggregation) using a variable number of parallel branches (data\_process).

\end{itemize}

\subsection{Benefits of Multiple FaaS in Joint$\lambda$}\label{section_benefits}

We first compare Joint$\lambda$ with ASF and AC on two real-world applications: video analytics and QA inference. Joint$\lambda$ deploys ResNet50 on Aliyun FC to accelerate inference by leveraging GPUs.


\begin{figure}[htbp]
    \centering
    \subfigure[P95 Makespan of video analytics workflow]{
    	\includegraphics[width=0.40\textwidth]{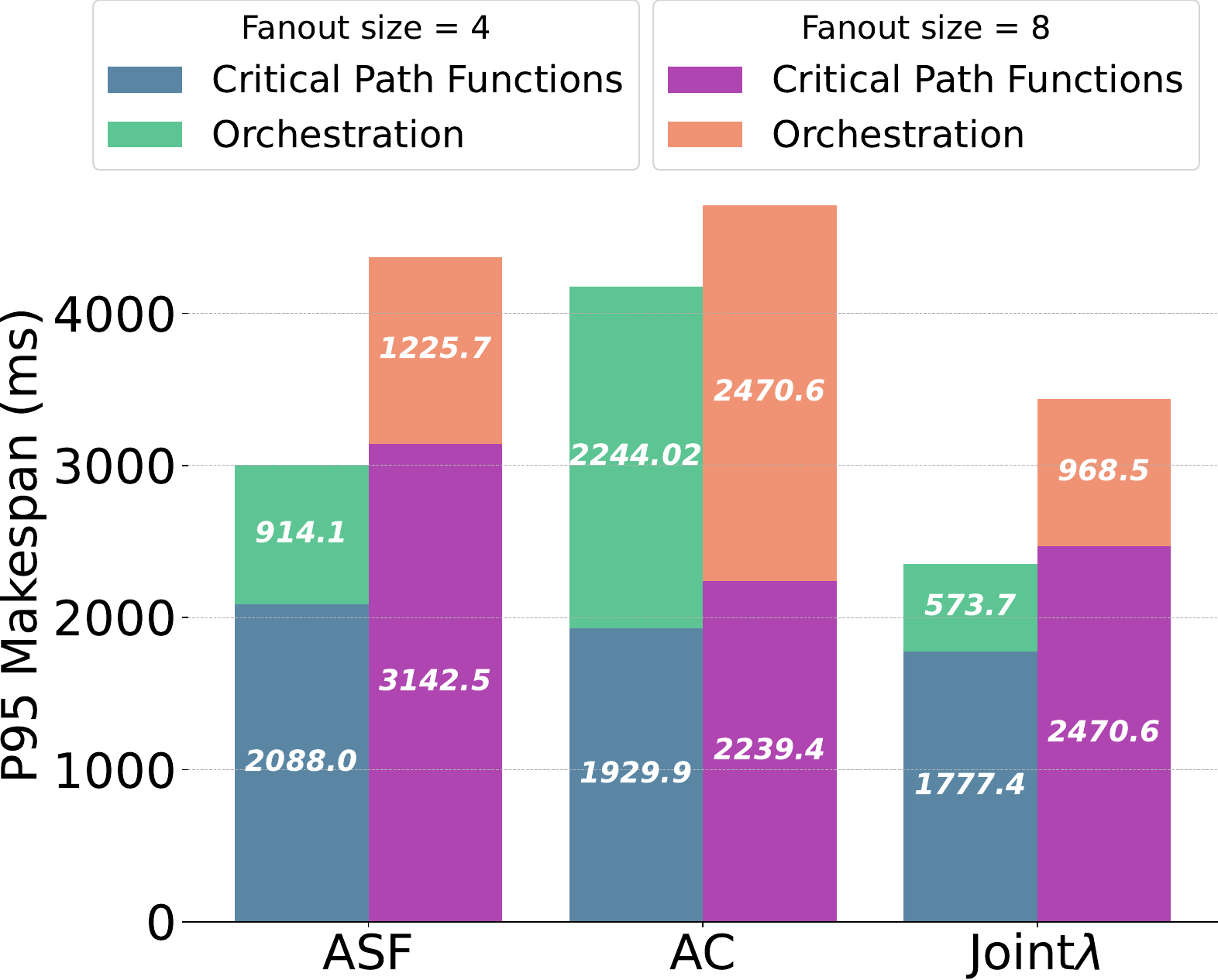}
    	\label{video_latency}
    	}
    \subfigure[Average cost of video analytics workflows when the fan-out size is 8]{
    	\includegraphics[width=0.40\textwidth]{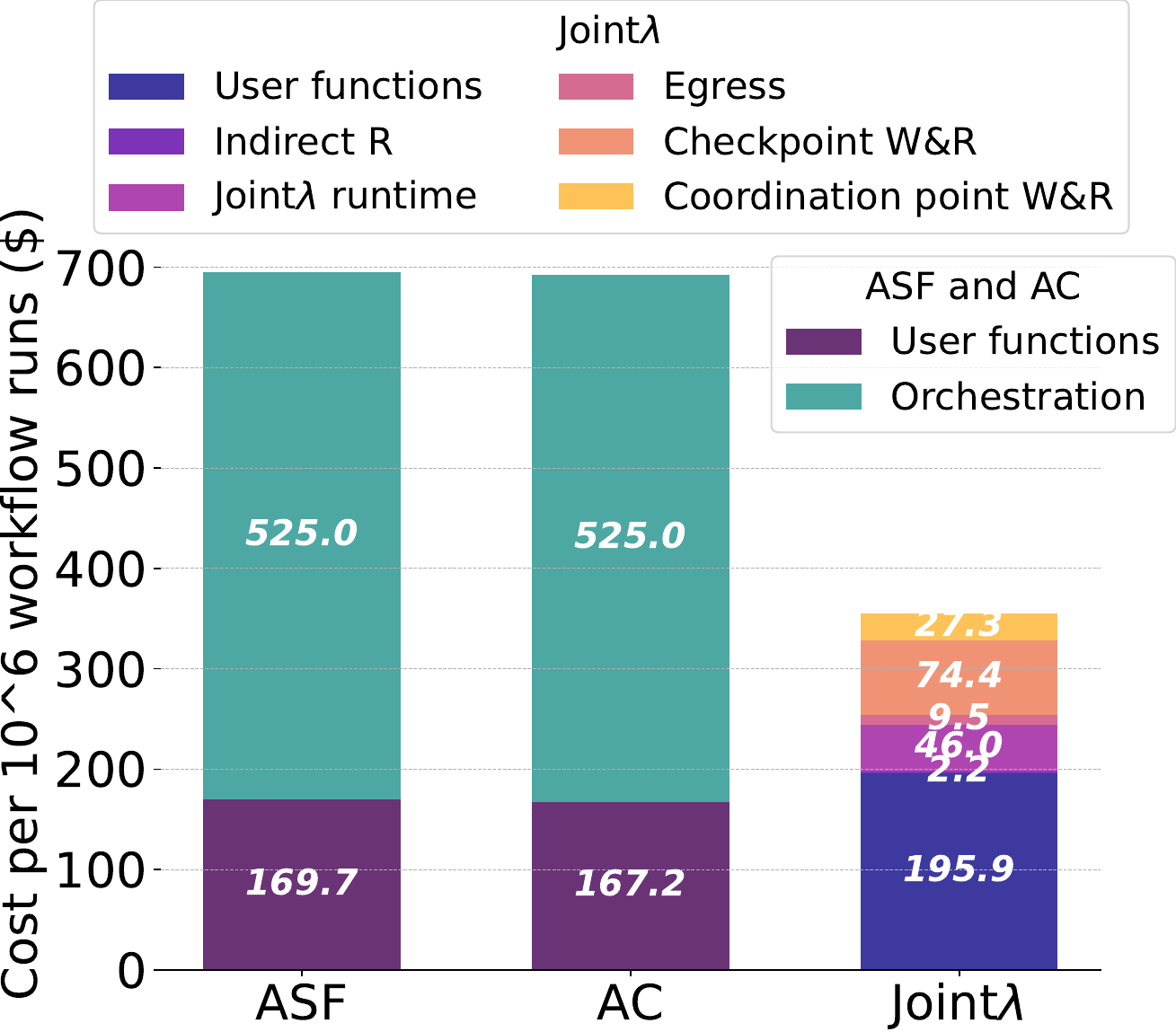}
    	\label{video_cost}
    	}
	\caption{The comparison between ASF, AC and Joint$\lambda$ implementation for Video Analytics workflow. Costs are computed per workflow run and linearly scaled to $10^6$ workflow runs for presentation.}
\end{figure}

Figure \ref{video_latency} displays the P95 end-to-end makespan achieved by all orchestrators. The Joint$\lambda$ implementation achieves the fastest execution, reducing makespan by 26\% and 21\% compared to AC and ASF, respectively, when the fan-out size is 8. Joint$\lambda$ also reduces makespan by 43\% and 21\% compared to AC and ASF, respectively, when the fan-out size is 4. Joint$\lambda$ consistently has lower orchestration overhead for both branch sizes due to its distributed orchestration design.
 
Figure \ref{video_cost} presents the total cost of video analytics workflow incurred by different orchestrators when the fan-out size is 8. Joint$\lambda$ saves at least 48\% cost compared to AC and ASF.
 Independent cloud orchestration
services charge \$25 per 1 million state transitions and the orchestration cost accounts for at least 75\%  of the total cost for both AWS  Step Functions and AC in this
workflow. However, only 44\% total cost is used for orchestration in Joint$\lambda$.  We observe that the cost of Checkpoint W\&R (write and read) takes up most of the orchestration cost, which is worth it because it can ensure fault tolerance. 

\begin{figure}[htbp]
    \centering
    \subfigure[P95 Makespan of QA workflow]{
    	\includegraphics[width=0.40\textwidth]{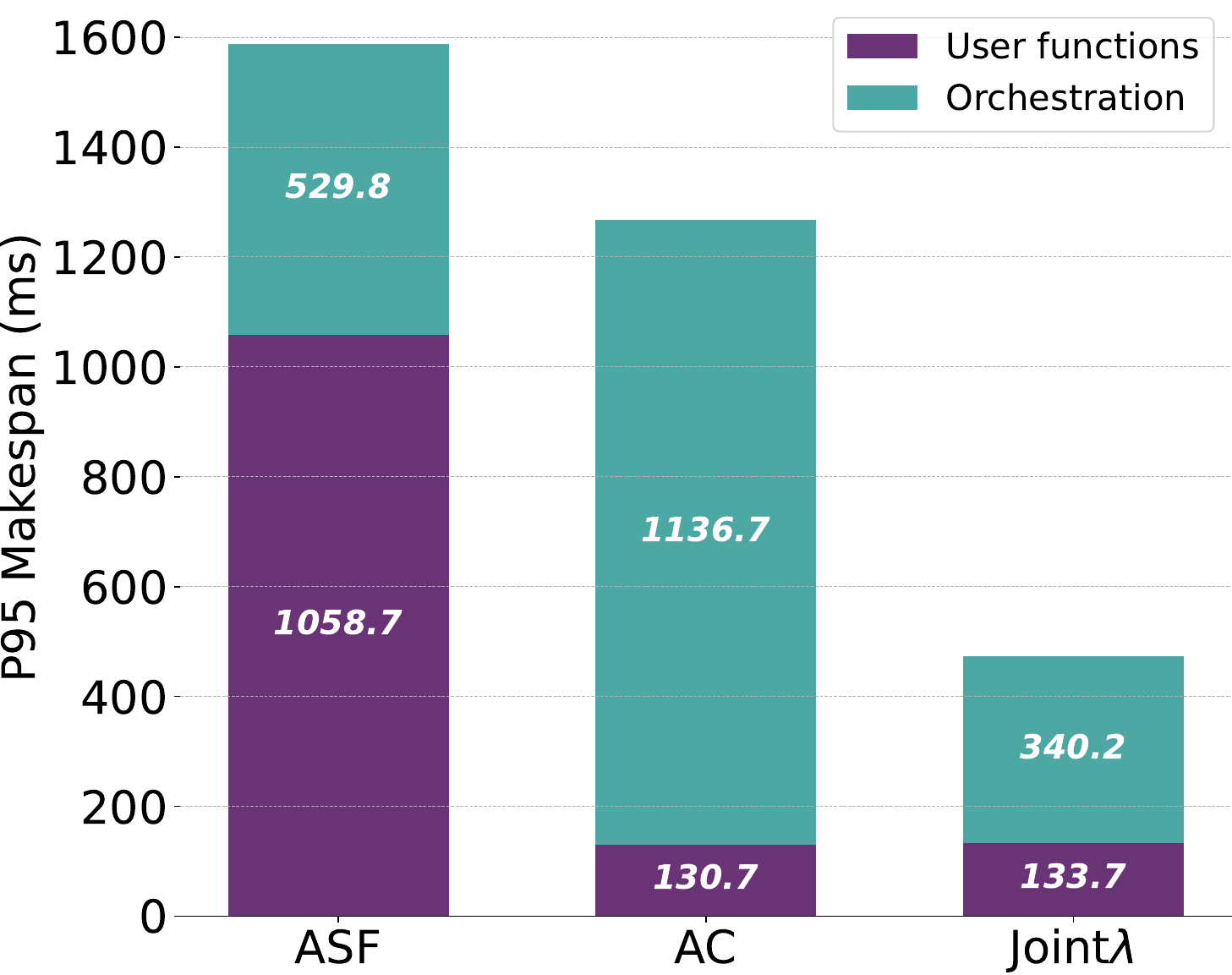}
    	\label{bert_latency}
    	}
    \subfigure[Average cost of QA workflow]{
    	\includegraphics[width=0.40\textwidth]{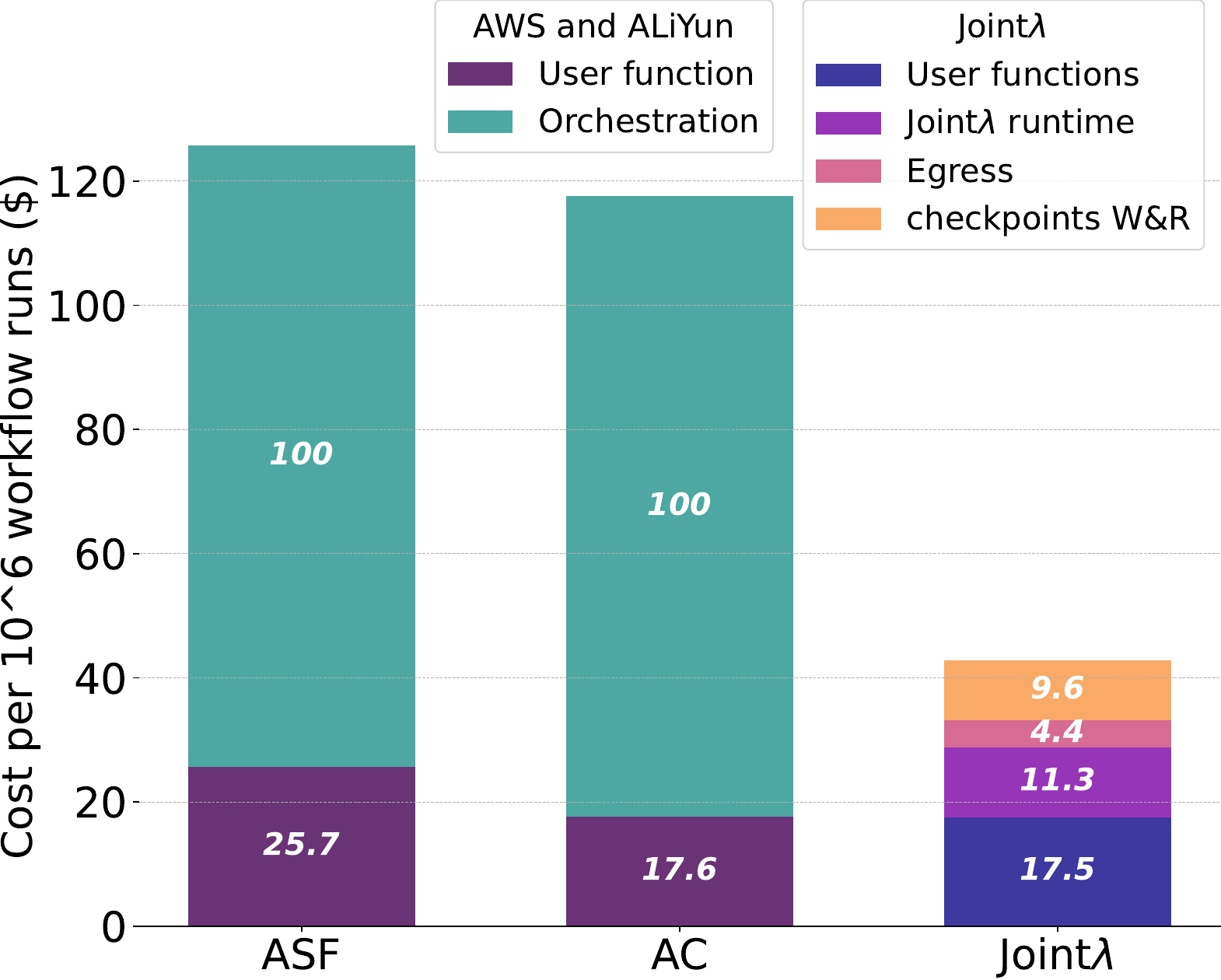}
    	\label{bert_cost}
    	}
	\caption{The comparison between ASF, AC and Joint$\lambda$ implementation for QA workflow. Costs are computed per workflow run and linearly scaled to $10^6$ workflow runs for presentation.}
\end{figure}

As shown in Figure \ref{bert_latency}, in the QA inference workflow, the Joint$\lambda$ implementation exhibits the fastest execution, delivering a 2.6$\times$ and 3.3$\times$ improvement over AC and ASF, respectively. The reduced workflow makespan mainly stems from inter-cloud heterogeneity and the difference in orchestration performance. Joint$\lambda$ fully utilizes heterogeneous accelerators in terms of user-functions computation. In terms of cost (Figure \ref{bert_cost}), Joint$\lambda$ achieves a 63\% cost savings over AC and a 65\% reduction compared to ASF. The majority of Joint$\lambda$'s orchestration cost (up to 26\%) is attributed to the Joint$\lambda$ runtime. The second largest cost factor is the checkpoint W\&R (Write and read), accounting for 22\% of the total orchestration expenditure.

To conclude, Joint$\lambda$ can leverage heterogeneous FaaS systems across clouds to accelerate functions execution thereby optimizing workflows.

\subsection{Failover Overhead}\label{section_recovery}


We use wrong invocations to inject failures in a controlled manner. Concretely, we intentionally configure a downstream function identifier that does not exist (e.g., an invalid function name on the target FaaS) in the MetaConfig, so that the invocation request deterministically fails and triggers Joint$\lambda$’s exception handling and failover logic. This emulates cloud-side unavailability from the orchestrator’s perspective without modifying the FaaS platform.

\begin{figure}[!htbp]
\centering
\includegraphics[width=0.40\textwidth]{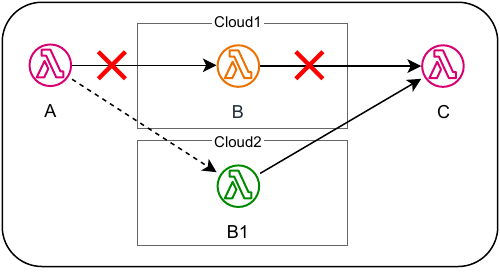}
\caption{Simulate cloud outages where a FaaS system is located.}
\label{ccf}
\end{figure}

 Figure \ref{ccf} shows the Joint$\lambda$ implementation (A, B, C, B1) and the single-cloud orchestrator implementation (A, B, C). Each function in the workflow executes a no operation instruction and is configured with 512 MB of RAM to ensure a fair comparison across systems. Joint$\lambda$ deploys function replica B1 on another FaaS platform in the same region, where the replica is launched only when invoked. For cross-cloud failover and invocation, the runtime must hold provider credentials or tokens that authorize calls to the target cloud. In our prototype, these credentials are provisioned to the runtime environment in advance; in practice, short-lived tokens or role-based credentials are preferable to long-lived static secrets to reduce security risk. We configure the workflow to be invoked every 100 ms and inject wrong invocations to simulate outages on the cloud platform hosting function B1 during the interval from the 10th to the 20th second after the start time.


\begin{figure}[!htbp]
    \centering
    \subfigure[Workflow on multiple FaaS based on Joint$\lambda$]{
\includegraphics[width=0.40\textwidth]{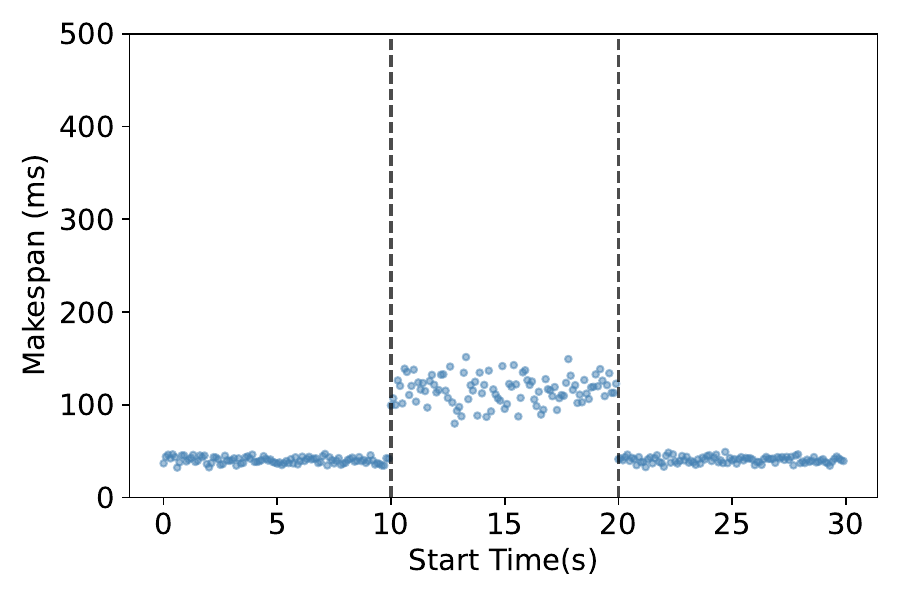}
    	\label{flt1}
    	}
    \hfill
    \subfigure[Workflow on single FaaS based on cloud orchestration services]{
\includegraphics[width=0.40\textwidth]{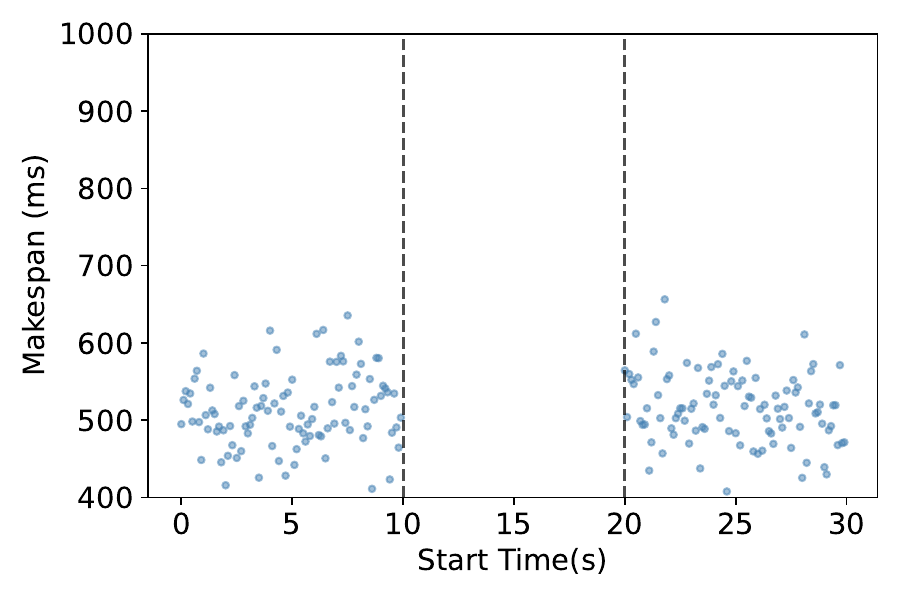}
    	\label{flt2}
    	}
	\caption{Makespan of workflows in the same region during cloud outage.}
    \label{mp}
\end{figure}

Figure \ref{mp} shows the end-to-end time of the workflow from 0 to 30 seconds. During the 10 to 20 second interval, the workflow on a single FaaS system exceeds the maximum number of retries for the orchestration service, resulting in a failure. It cannot execute successfully until the FaaS system recovers 20 seconds later, as shown in Figure \ref{flt2}. However, Figure \ref{flt1} shows that Joint$\lambda$ enables failover even after the failure of a FaaS system on a single cloud. It can automatically transfer the workflow to an alternate FaaS system in the same region to continue execution. 

In order to succinctly compare the time overhead of failover and normal execution, Figure \ref{flt1} shows the total execution time in both cases. The additional overhead introduced during failover averages about 78 ms. The increased makespan comes from creating alternative FaaS system clients and one additional cross-cloud invocation. Failover adds only 0.501\$ per 1M invocations in extra cost. Assuming the makespan SLO of the workflow is 300 ms, Joint$\lambda$ only violates the SLO during cold start. It reduces SLO violations by close to $99.9\%$ compared to a single-FaaS workflow and improves availability to nearly 1.


In conclusion, Joint$\lambda$ can efficiently disaster-transfer (failover) serverless workflows, incurring negligible additional overhead relative to the overall time and costs.

\subsection{Function-side Orchestration}\label{section_Optimization}

Besides the comparison with serverless workflows on a single FaaS, we also focus on the performance and cost overhead of Joint$\lambda$ and other state-of-the-art cross-platform orchestrators for orchestrating workflows on multiple FaaS systems. We choose a typical sequence workflow and a parallel-aggregate workflow to evaluate the overhead of orchestrating these basic workflow patterns. 


\begin{figure}[htbp]
    \centering
    \subfigure[P95 Makespan of IoT Pipeline for different number of functions]{
\includegraphics[width=0.40\textwidth]{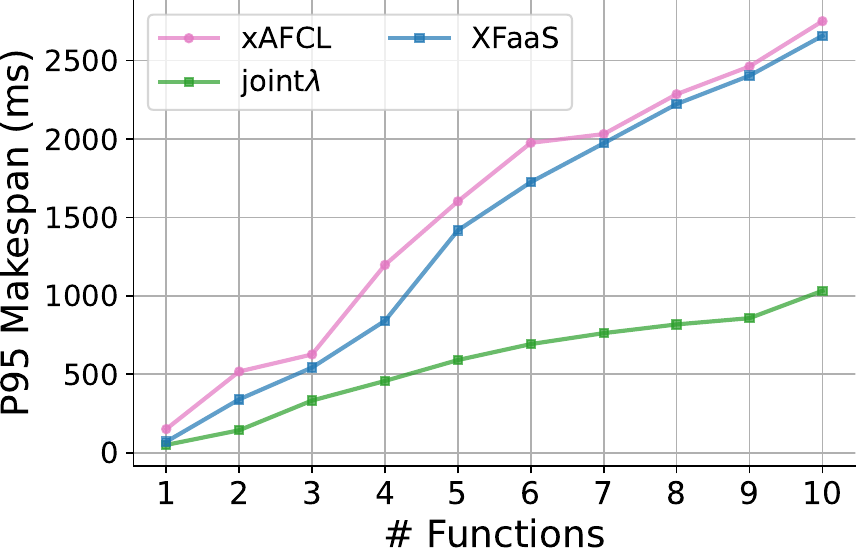}
    	\label{sequence_latency}
    	}
    \subfigure[P95 Makespan of MC workflows for different number of parallel branches]{\includegraphics[width=0.40\textwidth]{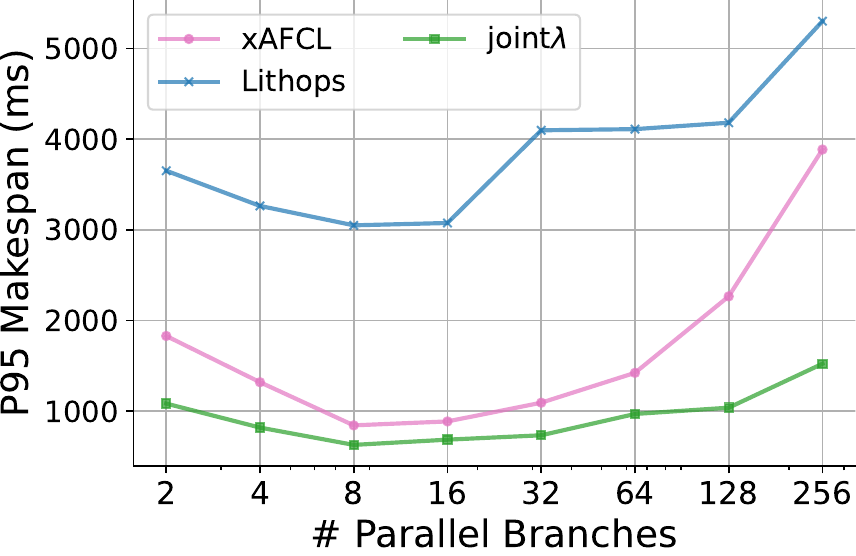}
    	\label{map_latency}
    	}
	\caption{P95 Makespan of IoT Pipeline and Monte Carlo Simulation.} \label{two_workflow_latency}
\end{figure}

\noindent \textbf{Sequence.}
We evaluate the orchestrator's state transition overhead using an IoT pipeline that passes data alternately on two FaaS systems. Since the execution time of the user function is very short (~10ms), the execution time of the workflow is essentially equal to the time of orchestration. Figure \ref{sequence_latency} shows the comparison of P95 makespan for Joint$\lambda$, xAFCL, and XFaaS for different lengths of IoT pipeline. For different numbers of functions, the total makespan of Joint$\lambda$ is consistently lower than that of xAFCL and XFaaS. At a length of 1, the makespans of the three orchestrators are close. As the number of functions increases, this gap between Joint$\lambda$ and the other two is gradually larger. This is due to increased cross-cloud transfers as the number of functions increases. When the number of functions is 10, Joint$\lambda$ is at least 2.5$\times$ faster than xAFCL and XFaaS. Compared to the other two centralized orchestrators, Joint$\lambda$ can improve orchestration performance by reducing cross-cloud transfer overhead.



\noindent \textbf{Parallel-aggregate.}
The parallel nature of MC simulation makes it an excellent candidate for our experiment. MC spawns a large number of parallel functions to partition computation. While XFaaS doesn't support fan-in orchestration, we compare Joint$\lambda$ with Lithops and xAFCL. The P95 makespan comparison of xAFCL, Lithops, and Joint$\lambda$ of MC workflow is shown in Figure \ref{map_latency}. When the number of parallel branches is 16, Joint$\lambda$ reduces the makespan by 22\% compared to xAFCL, and by 77\% compared to Lithops. As the number of parallel branches increases, the optimization results of Joint$\lambda$ become more significant compared to xAFCL and Lithops, improving 2.1$\times$ and 4.0$\times$ respectively with 128 parallel branches. 

Lithops' worker runtime initialization degrades its performance, introducing approximately 500 ms of additional makespan. xAFCL and Lithops are both limited by centralized orchestration, which incurs higher makespan as the number of branches increases. This indicates that the centralized bottleneck further limits parallel scalability. 

In short, Joint$\lambda$ achieves the best performance on these common workflow patterns. This result indicates that distributed control flow is beneficial for reducing workflow makespan when orchestrating workflows on Jointcloud FaaS systems.


\begin{table*}[htbp]
\caption{The cost comparison between Joint$\lambda$ and other orchestrators. N is the number of concurrent workflow runs. Totals are computed for $10^6$ workflow runs with $N{=}2$. For external orchestration VMs, we use m6g.2xlarge to reflect that orchestration is a lightweight long-running service and to avoid orchestration becoming the bottleneck. For VM-based datastores, we use m6g.large (0.099\$ per hour) as a modest configuration.}
\resizebox{\textwidth}{!}{ 
\begin{tabular}{c|c|c|c|c|c}
\hline
     Workflow & Orchestrator & Function Execution \& Invocation (cost) & External Orchestration & DataStore & Total (N=2) \\ \hline
\multirow{3}{*}{\makecell{IoT\\(Length: 10)}} & xAFCL &2.39\$ per 1M &0.396\$ per hour (m6g.2xlarge) &0.099\$ per hour (m6g.large) & 191.55\$ \\
 & XFaaS &4.86\$ per 1M &1500\$ per 1M &0\$ & 1504.86\$ \\
 & Joint$\lambda$ &5.95\$ per 1M &0\$ &48.50\$ per 1M (W\&R) & 54.45\$ \\ \hline
\multirow{4}{*}{\makecell{MC\\(fan-out size: 32)}} & xAFCL &11.12\$ per 1M &0.396\$ per hour (m6g.2xlarge) &0.099\$ per hour (m6g.large) & 86.17\$ \\
 & Lithops &59.64\$ per 1M & 0.396\$ per hour (m6g.2xlarge) &162.24\$ per 1M (W\&R) & 447.24\$ \\
 & Joint$\lambda$ &18.17\$ per 1M &0\$ &279.05\$ per 1M (W\&R) & 297.22\$ \\
 & Joint$\lambda$-VM &18.17\$ per 1M &0\$ &0.099\$ per hour (m6g.large) & 28.24\$ \\\hline                   
\end{tabular}}
\label{cost_compare}
\end{table*}

\noindent \textbf{Cost.}
The cost comparison between Joint$\lambda$ and the baselines is presented in Table~\ref{cost_compare}.
Joint$\lambda$ achieves on-demand billing for serverless workflows on Jointcloud FaaS systems, charging only for the resources actually used.
However, most existing cross-platform orchestrators require VMs to host orchestration and datastore nodes. These VMs are billed by the hour, and their total cost depends on runtime duration and resource utilization.
For a relatively fair comparison among VM-hosted baselines, we assume that there are $N$ concurrent workflow runs (i.e., $N$ workflows execute in parallel) and that the utilization rate of the VM is 100\%.
Thus the cost of the VM can be estimated as $(unit\_price * M * T)/N$, where $unit\_price$ is the price of the VM per hour, $M$ is the number of workflow invocations (e.g., $10^6$), and $T$ is the end-to-end time (makespan) of the workflow, as shown in Figure~\ref{two_workflow_latency}.
We conservatively use the higher price between Amazon DynamoDB and Aliyun TableStore, i.e., 1.4269\$ per 1M writes and 0.285\$ per 1M reads.
``Function Execution \& Invocation'' is computed from the measured average function duration (under its configured memory) using the providers' on-demand billing rates plus the per-request charge, and then scaled to $10^6$ workflow runs. For Joint$\lambda$'s managed datastore cost (W\&R), we count protocol-induced reads and writes and multiply them by the per-million request prices, yielding 48.50\$ for the IoT workflow and 279.05\$ for MC (fan-out~=~32). For Lithops, the managed datastore cost is computed from per-request pricing of the underlying object-storage operations, resulting in 162.24\$ for MC (fan-out~=~32). For XFaaS, we model the external orchestration overhead using transition-based billing of managed workflow services: following the DAG in Figure~\ref{workload}, an IoT pipeline of length~=~10 incurs about 20 orchestration steps, and each hop involves three service-level state transitions, resulting in $3 \times 20{=}60$ transitions per workflow run. Under a \$0.025 per 1000-transitions price, this gives $60 \times 10^6 / 1000 \times 0.025{=}\$1500$ per $10^6$ runs.
Since the egress fees for cross-cloud workflows under multiple orchestrators are very close, we do not discuss them.

Joint$\lambda$ is more expensive in terms of function execution cost because each function needs to run additional runtime libraries. Lithops is also expensive because its worker runtime initialization time is longer. However, Joint$\lambda$ does not require a standalone orchestrator, so its external orchestration cost is 0. For data storage, Joint$\lambda$ uses managed storage services and therefore incurs charges for checkpoints and coordination points.
For the IoT workflow with 10 functions, Joint$\lambda$ is about 3.5$\times$ and 27.6$\times$ cheaper than xAFCL and XFaaS, respectively.
For the MC workflow (fan-out size = 32), xAFCL becomes cheaper when provisioned with a modest external orchestration VM, while Joint$\lambda$ remains 1.5$\times$ cheaper than Lithops. Joint$\lambda$ provides execution guarantees for a large number of parallel functions, which increases datastore cost. Based on this analysis, we can rent a VM to host the datastore node to reduce the cost, as shown by Joint$\lambda$-VM in Table~\ref{cost_compare}, which further reduces the total cost by about 3.1$\times$ compared to xAFCL.



\subsection{Orchestration Overhead of Joint$\lambda$}\label{section_Overhead}

\begin{figure}[!htbp]
    \centering \includegraphics[width=0.4\textwidth]{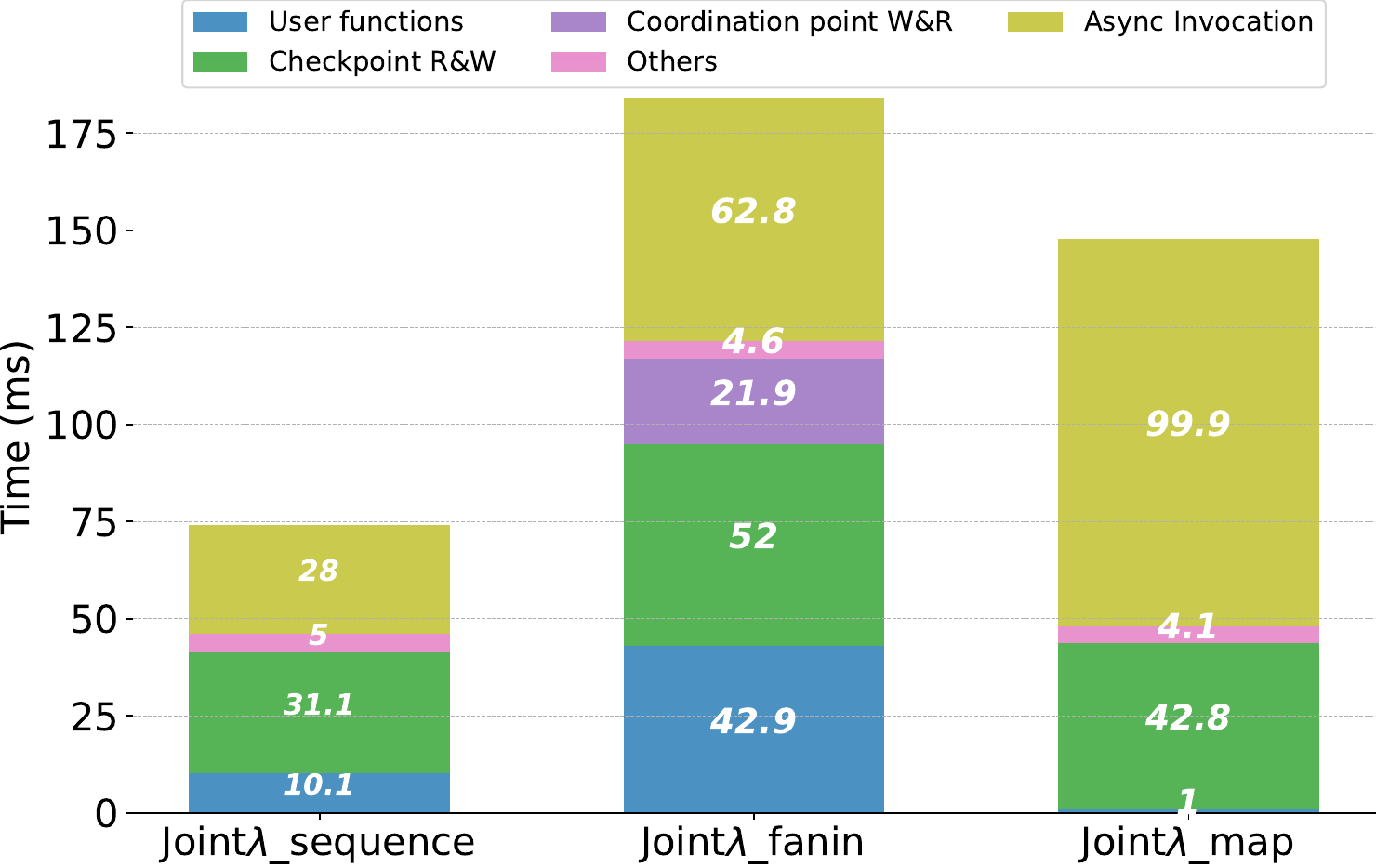}
	\caption{Joint$\lambda$'s cross-cloud sequence, fan-in and map overhead. Waiting time for asynchronous queue after invocation is not added.} \label{fan_out_in_overhead}
\end{figure}

To probe the latency introduced by the function-side orchestrator of Joint$\lambda$, Figure~\ref{fan_out_in_overhead} presents decomposed latency traces for a sequence function (from AWS Lambda to Aliyun FC), a fan-in function, and map functions from the IoT pipeline and MC simulation (fan-out size = 32). 

For sequence invocation mode, most of the overhead is writing and reading checkpoints, which take up 48.5\% of the total Joint$\lambda$ runtime. Each record function execution introduces 3W2R (3 writes and 2 reads) datastore operations, where 2W1R are for invocation checkpoints and 1W1R are for output data checkpoints. 

For map invocation mode, asynchronous invocation is the largest overhead, accounting for 68\% of the total Joint$\lambda$ runtime. The data\_map function invokes 32 functions concurrently, resulting in higher asynchronous invocation overhead. Since the number of subsequent (downstream) functions is greater than 10, Joint$\lambda$ writes invocation checkpoints in batches, i.e., it performs batch writes 4 times. Each execution of the function generates 5W1R datastore operations for the invocation checkpoint, while the output data checkpoint still generates 1W1R. The performance bottleneck of data\_map lies in the numerous cross-cloud parallel invocations, and the use of batched checkpoint writes effectively reduces the time overhead. 
The data\_process function, which uses fan-in invocation mode, increases execution time due to the additional operations required for reading and writing the bitmap (coordination point W\&R). The coordination point additionally introduces 2W2R datastore operations. The last completed parallel data\_process function invokes an aggregate function. Each function execution also introduces 3W2R datastore operations. The differences between it and the record function in terms of checkpoint latency and invocation latency stem from the heterogeneity of cloud platforms.

In summary, Joint$\lambda$ introduces modest orchestration latency. The main performance overhead is in reading and writing the datastore and in a large number of parallel function invocations.


\section{Conclusion}
This paper proposes Joint$\lambda$, a distributed runtime system for orchestrating serverless workflows across multiple FaaS systems. Joint$\lambda$ leverages inter-cloud heterogeneity to reduce completion time and cost. By using function-side orchestration instead of centralized nodes, it enables independent function invocations and data transfers, thereby reducing cross-cloud communication overhead. In addition, Joint$\lambda$ supports exactly-once execution semantics and provides fault tolerance and failover in multi-cloud environments. Joint$\lambda$ allows users to flexibly design workflows that exploit inter-cloud heterogeneity. Compared with standalone cloud providers' orchestrators and state-of-the-art cross-cloud serverless workflow engines, Joint$\lambda$ consistently improves performance for cross-cloud workflow orchestration and can reduce cost in representative settings.

\section*{Acknowledgements}
This work is supported by the National Natural Science Foundation of China (No.~62202479 and No.~61772030) and the National Key Laboratory Funding Project (No.~2025-KJWPDL-04). During the preparation of this work, the authors used Gemini\cite{google_gemini} to polish the manuscript. After using this service, the authors reviewed and edited the content as needed and take full responsibility for the content of the published article.
\bibliography{ref2bib/references}
\end{document}